\documentclass[twocolumn,twocolappendix]{aastex631}

\usepackage{multirow}
\usepackage{gensymb}
\newcommand{\hbeta}{H{$\beta$}}
\newcommand{\halpha}{H{$\alpha$}}

\newcommand{\OI}{[O{\sevenrm\,I}]}

\newcommand{\OIII}{[O{\sevenrm\,III}]}
\newcommand{\loiii}{$L_{\text{[O \textrm{\tiny III}]}}$}

\newcommand{\OIIIb}{[O{\sevenrm\,III}]\,$\lambda$5007}

\newcommand{\NIIb}{[N{\sevenrm\,II}]\,$\lambda$6584}

\newcommand{\SIIab}{[S{\sevenrm\,II}]\,$\lambda\lambda$6717,~6731}
 
 \font\sevenrm=cmr7 scaled 1000

\newcommand{\statm}{\texttt{statmorph}}
\newcommand{\acas}{\mbox{$A_\mathrm{CAS}$}}
\newcommand{\aout}{\mbox{$A_\mathrm{outer}$}}
\newcommand{\asha}{\mbox{$A_\mathrm{shape}$}}
\newcommand{\abkg}{\mbox{$A_\mathrm{bkg}$}}
\newcommand{\rpet}{\mbox{$R_\mathrm{P}$}}

\shorttitle{Asymmetry in Low-redshift Galaxies}
\shortauthors{Zhao et al.}
\graphicspath{{./}{figures/}}

\begin{document}

\title{The Relation between Morphological Asymmetry and Nuclear Activity in Low-redshift Galaxies}
\correspondingauthor{Jinyi Shangguan}
\email{shangguan@mpe.mpg.de}

\author[0000-0003-4591-2532]{Yulin Zhao}
\affil{Kavli Institute for Astronomy and Astrophysics, Peking University, Beijing 100871, China}
\affiliation{Department of Astronomy, School of Physics, Peking University, Beijing 100871, China}

\author[0000-0002-3309-8433]{Yang A. Li}
\affil{Kavli Institute for Astronomy and Astrophysics, Peking University, Beijing 100871, China}
\affiliation{Department of Astronomy, School of Physics, Peking University, Beijing 100871, China}

\author[0000-0002-4569-9009]{Jinyi Shangguan}
\affil{Max-Planck-Institut f\"{u}r Extraterrestrische Physik (MPE), Giessenbachstr., D-85748 Garching, Germany}
\affiliation{Kavli Institute for Astronomy and Astrophysics, Peking University, Beijing 100871, China}

\author[0000-0001-5105-2837]{Ming-Yang Zhuang}
\affil{Kavli Institute for Astronomy and Astrophysics, Peking University, Beijing 100871, China}
\affiliation{Department of Astronomy, School of Physics, Peking University, Beijing 100871, China}

\author[0000-0001-6947-5846]{Luis C. Ho}
\affil{Kavli Institute for Astronomy and Astrophysics, Peking University, Beijing 100871, China}
\affiliation{Department of Astronomy, School of Physics, Peking University, Beijing 100871, China}

\begin{abstract}
The morphology of galaxies reflects their assembly history and ongoing dynamical perturbations from the environment.  Analyzing stacked $i$-band images from the Pan-STARRS1 $3\pi$ Steradian Survey, we study the optical morphological asymmetry of the host galaxies of a large, well-defined sample of nearby active galactic nuclei (AGNs) to investigate the role of mergers and interactions in triggering nuclear activity. The AGNs, comprising 245 type~1 and 4514 type~2 objects, are compared with 4537 star-forming galaxies matched in redshift ($0.04<z<0.15$) and stellar mass ($M_*>10^{10}\,M_\odot$).  We develop a comprehensive masking strategy to isolate the emission of the target from foreground stars and other contaminating nearby sources, all the while retaining projected companions of comparable brightness that may be major mergers.  Among three variants of nonparametric indices, both the popular CAS asymmetry parameter (\acas) and the outer asymmetry parameter (\aout) yield robust measures of morphological distortion for star-forming galaxies and type~2 AGNs, while only \aout\ is effective for type~1 AGNs.  The shape asymmetry ($A_{\rm shape}$), by comparison, is affected more adversely by background noise.  Asymmetry indices $\gtrsim 0.4$ effectively trace systems that are candidate ongoing mergers. Contrary to theoretical expectations, galaxy interactions and mergers are not the main drivers of nuclear activity, at least not in our sample of low-redshift, relatively low-luminosity AGNs, whose host galaxies are actually significantly less asymmetric than the control sample of star-forming galaxies.  Moreover, type~2 AGNs are morphologically indistinguishable from their type~1 counterparts. The level of AGN activity does not correlate with asymmetry, not even among the major merger candidates.  As a by-product, we find, consistent with previous studies, that the average asymmetry of star-forming galaxies increases above the main sequence, although not all major mergers exhibit enhanced star formation.
\end{abstract}

\keywords{galaxies: evolution --- galaxies: formation --- galaxies: active --- galaxies: bulges --- galaxies: photometry --- quasars: general}

\section{Introduction} \label{sec1:intro}

Supermassive black holes (BHs) in the centers of galaxies grow mainly through the accretion of gas \citep{Soltan1982MNRAS}, which loses angular momentum most effectively through gravitational torques imparted by dynamical interactions and mergers with neighboring galaxies (e.g., \citealt{Hernquist1989Natur,Barnes1992ARAA,DiMatteo2005Natur,Springel2005MNRAS}).  This basic theoretical expectation has led to a long historical prejudice that galaxies hosting active galactic nuclei (AGNs) should manifest signatures of dynamical perturbation, most readily in the form of external, large-scale morphological asymmetry, such as tidal tails, lopsidedness, and perhaps even direct evidence of ongoing disturbance by a visible companion.  Moreover, if AGNs are triggered externally by dynamical perturbations, one would naively expect a correlation between the level of nuclear activity and the degree of perturbation.

Previous investigations paint a confusing picture.  While many studies argue that both the incidence and intensity of nuclear activity are linked to galaxy mergers and interactions (e.g., \citealt{Ellison2011MNRAS,Silverman2011ApJ,Bessiere2012MNRAS,Treister2012ApJ,Hong2015ApJ,Goulding2018PASJ,Kim2021ApJS}), the evidence becomes much murkier for less luminous AGNs (e.g., \citealt{Grogin2005ApJ,Cisternas2011ApJ,Villforth2014MNRAS,Villforth2017MNRAS}).  X-ray-selected AGNs of moderate luminosity are especially challenging because the majority reside in relatively late-type galaxies, whose very disky nature casts doubt on the viability of the merger scenario for triggering the activity \citep{Schawinski2011ApJ,Schawinski2012MNRAS,Kocevski2012ApJ,Simmons2012ApJ}. The resolution to these seemingly conflicting results lies in the realization that the triggering mechanism depends on the level of AGN activity \citep{Hopkins2009ApJ,Draper2012ApJ}.  Whereas the most powerful quasars require a prodigious accretion rate that can only be supplied by an external event as a galaxy merger, run-of-the-mill Seyferts, not to mention of the even lower luminosity, radiatively inefficient nearby nuclei, can be sustained by internal secular processes \citep{Ho2008ARAA,Ho2009ApJ}.

A closely related topic concerns the physical connection between AGNs of different types. While the orientation-based unified model \citep{Antonucci1993ARAA,Urry1995PASP} has enjoyed much success in explaining the relationship between broad-line (type~1) and narrow-line (type~2) AGNs, there has been mounting observational evidence that the two AGN types must possess true intrinsic differences, which are in part or in whole mediated by evolution (e.g., \citealt{Sanders1988ApJ}).  An incomplete list includes differences in host galaxy morphology, star formation rate (SFR), velocity field, and environment \citep{Maiolino1997ApJ,Kim2006ApJ,Lacy2007ApJl,Greene2011ApJ,Villarroel2014NatPh,Chen2015ApJ,Villarroel2017ApJ,Zhuang2020ApJ}.

Several practical hurdles need to be overcome in designing an effective experiment to test the aforementioned ideas.  Many older studies were plagued by small-number statistics, which, fortunately, can now be ameliorated using modern large-area surveys.  Constructing an appropriate control sample is also critically important to achieving a meaningful statistical comparison for hypothesis testing.  This, too, is helped enormously with the availability of survey data.  Finally, one should endeavor to abandon subjective visual inspection and simplistic qualitative assessment \citep{Lambrides2021ApJ}, and strive toward securing more rigorous, objective metrics to quantify the morphological properties of galaxies.

The primary goal of this paper is to re-evaluate the question of whether the host galaxies of AGNs are preferentially more morphologically disturbed than a control sample of inactive galaxies of otherwise similar general makeup.  As a corollary, we evaluate whether the level of AGN activity varies with the degree of morphological disturbance.  Comprising both type~1 and type~2 sources, our AGN sample can further test for potential differences between the two types, as might arise if they evolve from one to the other.

We take advantage of nonparametric measurements of galaxy structure, with emphasis on several parameters commonly used to quantify the degree of morphological asymmetry \citep{Conselice2003ApJS, Wen2014ApJ, Pawlik2016MNRAS}, and apply them to high-quality ground-based images from Data Release 2 of the Pan-STARRS1 $3\pi$ Steradian Survey \citep{Chambers2016,Flewelling2020ApJS}.  The stacked images of PanSTARRS have reasonable sensitivity ($5\,\sigma$ depth of 23.3, 23.2, 23.1, 22.3, and 21.3 mag in the $g$, $r$, $i$, $z$, and $y$ band) and resolution [median full-width at half-maximum (FWHM) of the point-spread function (PSF) is 1\farcs31, 1\farcs19, 1\farcs11, 1\farcs07, and 1\farcs02, respectively].  The original targets were selected from the Sloan Digital Sky Survey (SDSS; \citealt{York2000AJ}) value-added MPA-JHU catalog \citep{Brinchmann2004MNRAS} of star-forming galaxies (SFGs) and type~2 AGNs (AGN2s), supplemented by type~1 AGNs (AGN1s), which are not included in the MPA-JHU database, from the catalog of \cite{Liu2019ApJS}.  

We carefully match the SFG, AGN2, and AGN1 samples according to their stellar mass and redshift (Section~\ref{sec2:data}).  After developing a comprehensive strategy to generate an effective mask to isolate the targets of interest from contaminating sources (Section~2.3), we employ the \statm\ package \citep{Rodriguez-Gomez2018MNRAS} to calculate three frequently used nonparametric measures of source asymmetry (Section~\ref{sec4:asy}): the CAS asymmetry (\acas; \citealt{Abraham1994ApJ,Abraham1996ApJS,Conselice2000ApJ,Conselice2003ApJS}), the outer asymmetry (\aout; \citealt{Wen2014ApJ}), and the shape asymmetry (\asha; \citealt{Pawlik2016MNRAS}).  We statistically compare the asymmetries of the three galaxy samples, assess the incidence of major mergers, and discuss the connection between asymmetry and nuclear activity and star formation (Section~\ref{sec5:res}).  A summary is given in Section~5.  This paper assumes a \cite{Chabrier2003PASP} stellar initial mass function and a cosmology with $H_0=70$ km~s$^{-1}$~Mpc$^{-1}$, $\Omega_m=0.3$, and $\Omega_{\Lambda}=0.7$.

\section{Sample and Data Analysis} \label{sec2:data}

\subsection{Sample Selection} \label{sec2:data}

We restrict our sample to low redshift ($0.04<z<0.15$) in order to avoid any significant effect from cosmological evolution.  We require $z>0.04$ to ensure that the 3\arcsec\ fiber of SDSS covers at least $20\%$ of each galaxy, to mitigate aperture effects on emission-line measurements \citep{Kewley2005PASP}.  After extensive visual inspection and experimentation, we find that the moderate spatial resolution of the Pan-STARRS images renders it difficult to identify clear morphological features for typical galaxies at $z\gtrsim 0.2$.  This is consistent with the study of \cite{Ferreira2018MNRAS}, who conclude that morphological classification of galaxies becomes challenging for $z \gtrsim 0.2$ in images with PSF ${\rm FWHM} \gtrsim 1\arcsec$.  We therefore limit our sample redshift to $z<0.15$.  

The MPA-JHU DR7 catalog \citep{Brinchmann2004MNRAS} provides the emission-line measurements of active and inactive alaxies with narrow emission lines.  Following \cite{Kewley2006MNRAS}, we adopt the emission-line intensity ratio diagnostic diagrams of \cite{Baldwin1981PASP} to classify objects as star-forming galaxies (SFGs), Seyferts, low-ionization nuclear emission-line regions (LINERs), and composite sources. To obtain a robust classification, we  require the signal-to-noise ratio (S/N) of at least 3 for \halpha, \hbeta, \OIIIb, \NIIb, and \SIIab.  No S/N constraint was placed on the \OI~$\lambda 6300$ line, as it is typically very weak.  For the purposes of this study, our sample of AGN2s only considers objects classified as Seyferts, whose accretion luminosities most closely resemble those of the comparison sample of AGN1s selected from \cite{Liu2019ApJS}.  We omit LINERs and composite sources, which, apart from  having characteristically lower luminosities, lower Eddington ratios, and a distinctly different accretion mode compared to the Seyferts \citep{Ho2009ApJ}, also suffer most readily from confusion by non-nuclear sources of excitation in large-aperture spectra such as those of SDSS \citep{Ho2008ARAA}.

As the majority of nearby, optically selected AGNs reside in relatively massive galaxies \citep{Ho1997ApJ,Kauffmann2003MNRAS}, we choose a stellar mass cut of $M_* > 10^{10}\,M_\odot$.  We take advantage of the GALEX-SDSS-WISE Legacy Catalog 2 (GSWLC-2; \citealt{Salim2018ApJ}), which provides total stellar masses and SFRs based on analysis of the spectral energy distribution covering photometric measurements of the entire galaxy from the ultraviolet to the mid-infrared bands.  These measurements are applicable to the SFG and AGN2 subsamples.  For the AGN1 sources, whose unobscured active nucleus pollutes the integrated magnitude and colors of the host galaxy, we derive $M_*$ from the empirical relation between BH mass and total stellar mass, as recently calibrated by \cite{Greene2020ARAA}.\footnote{We use the relation for all galaxy types, $\log\,M_* = (\log\,M_\mathrm{BH}-7.43)/1.61+10.477$.}  Although uncertain (0.81~dex intrinsic scatter), this method provides practical, unbiased estimates of the stellar mass of AGN1s for statistical study.  The BH masses,  with uncertainty 0.3--0.5~dex, were derived by \cite{Liu2019ApJS} using the \hbeta\ virial BH mass estimator (for all bulge types) of \cite{Ho2015ApJ} and the conversion between \halpha\ and \hbeta\ line width from \cite{Greene2005ApJ}.

Since the fraction of galaxy mergers depends on the mass of the system \citep{Hopkins2010ApJ}, we match the stellar mass of the three samples.  We further match the redshift distribution of the samples because the galaxy size relative to the spatial resolution also affects asymmetry measurements \citep{Thorp2021MNRAS}.  Given the large uncertainty of the individual stellar masses for AGN1s, we first match the distribution of redshifts and stellar masses for the SFGs and AGN2s, and then proceed to match the AGN1s to the distributions of SFGs and AGN2s.  The sample matching is done following the acceptance-rejection method used in \cite{Zhuang2020ApJ}, in redshift bins of size 0.02 and stellar mass bins of size of size 0.1~dex, which yield an adequate number of objects per bin to perform a Kolmogorov-Smirnov (K-S) test, which is used extensively in our later analysis (Section~\ref{sec5:res}).
 
\begin{figure}
\centering
\includegraphics[width=0.45\textwidth]{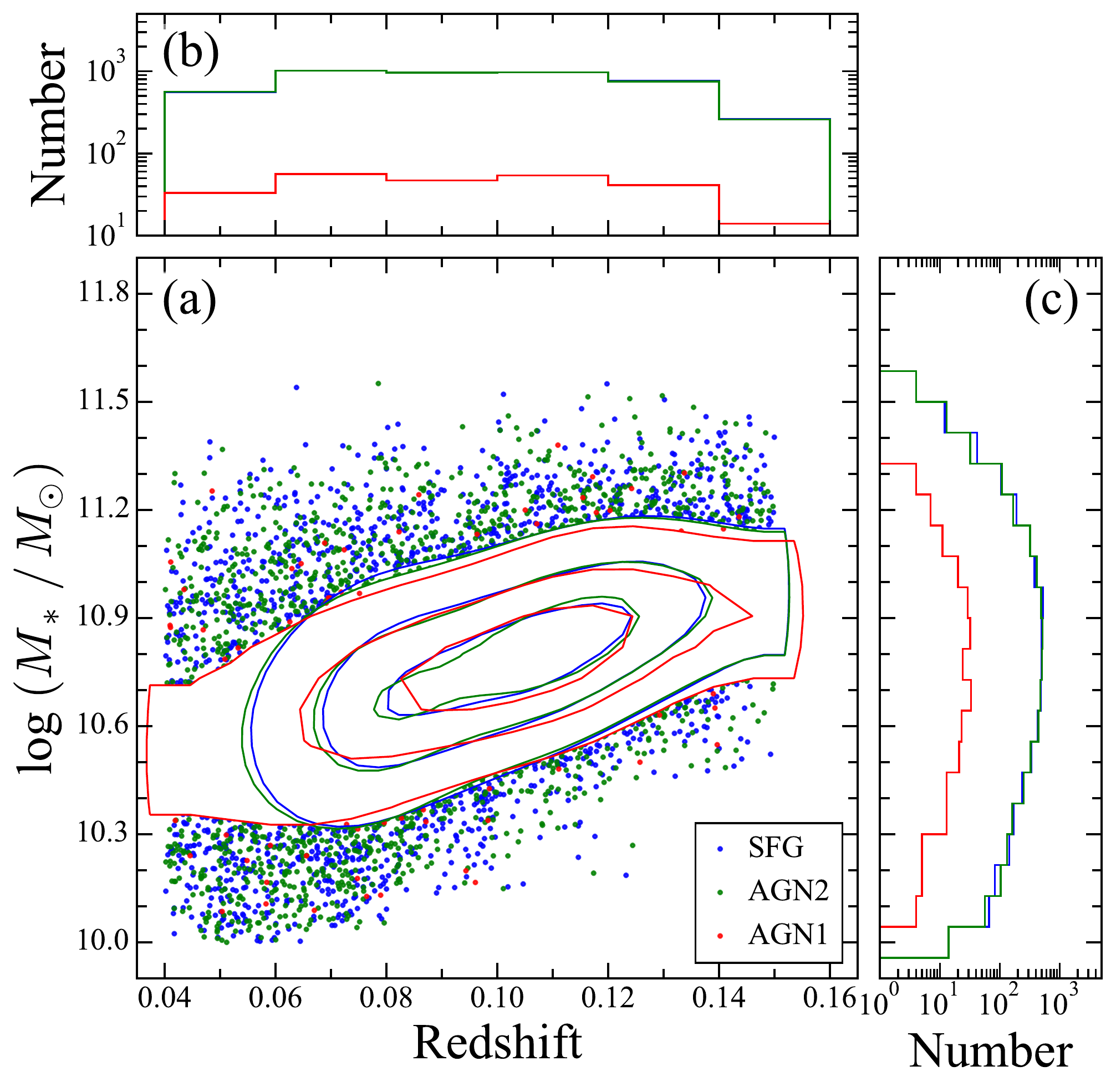}
\caption{ (a) The final sample of 4537 star-forming galaxies (SFGs; blue), 4514 type~2 AGNs (AGN2; red), and 245 type~1 AGNs (AGN1; green) matched in (b) redshift and (c) stellar mass.  For clarity, the sample distributions in the dense regions are represented by contours, with levels of 20 and 40 percentile.}
\label{Fig:final_sample}
\end{figure}

The MPA-JHU catalog (see, e.g., \citealt{Kauffmann2003MNRAS}) includes some so-called intermediate-type (type~1.9) AGNs, which exhibit very weak broad \halpha\ but no broad \hbeta\ emission \citep{Osterbrock1989,Ho1997ApJS}. We regroup 22 such objects as AGN1s.  After removing targets without robust asymmetry measurements (Section~\ref{sec4:asy}), our final sample, summarized in Table~\ref{tab:result},  consists of 4537 SFGs, 4514 AGN2s, and 245 AGN1s.  The three subsamples are well-matched in stellar mass and redshift according to the K-S test (Figure~\ref{Fig:final_sample}).  The small number of AGN1s relative to AGN2s in our final sample is a consequence of our mass-matching requirement.  In a flux-limited survey, selection effects will bias AGN1s to fainter (lower mass) host galaxies than AGN2s because the total flux of AGN1s includes a nonstellar contribution from the bright active nucleus.

\begin{figure*}[htbp]
\centering
\includegraphics[width=0.9\textwidth]{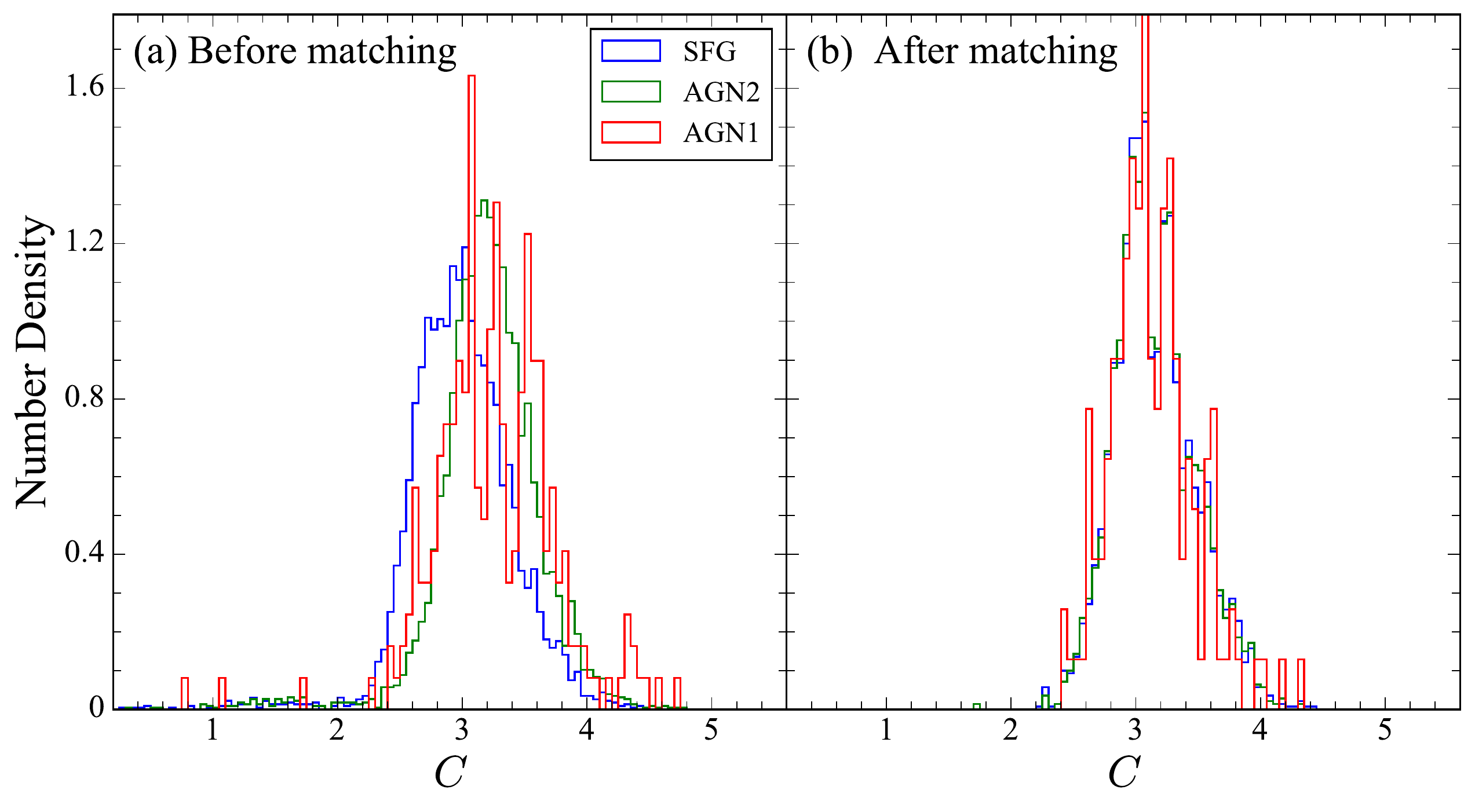}
\caption{Distribution of optical concentration ($C$) for the star-forming galaxy sample (SFGs; blue), the type~2 AGN sample (AGN2; green), and the type~1 AGN sample (AGN1; red), (a) before and (b) after matching their distributions.}
\label{Fig:C_match}
\end{figure*}

Galaxies of a given stellar mass can have different bulge-to-total ratio or light concentration.  Following standard practice \citep{Bershady2000AJ,Conselice2003ApJS}, we define the concentration parameter as $C=5\log\,(R_{80}/R_{20})$, where $R_{80}$ is the radius that contains 80\% of the light within 1.5 times the \cite{Petrosian1976ApJ} radius $\rpet$, and $R_{20}$ is the radius that contains 20\% of the light within $1.5 \,\rpet$. Figure~\ref{Fig:C_match}a illustrates that SFGs have lower $C$ (mean and standard deviation $2.99\pm0.42$) than AGN1s ($3.24\pm0.49$) or AGN2s ($3.21\pm0.41$) according to the Student's $t$-test ($p <10^{-3}$), while both AGN types are statistically similar ($p = 0.11$).  This is unsurprising, for AGNs are selected to contain accreting central BHs, most \citep{Kormendy&Ho2013}, if not all \citep{Greene2020ARAA}, of which reside in bulges. Low-redshift AGNs are preferentially hosted by earlier type galaxies (\citealt{Ho1997ApJ,Kauffmann2003MNRAS,Bruce2016MNRAS,Kim2017ApJS,Kim2021ApJS}) which tend to have more prominent bulges \citep{Simien1986ApJ,Gao2019ApJS} and hence higher concentration \citep{Strateva2001AJ}. Moreover, earlier type galaxies tend to have lower asymmetry \citep[e.g.,][]{Conselice2003ApJS}. To ensure a rigorous match between the active and inactive samples, it is desirable to control for the concentration.  In the subsequent analysis, we study the statistics of the asymmetry parameters using the samples with and without matching for $C$ (Figure~\ref{Fig:C_match}).  However, we find that our main conclusions are not sensitive to this choice, and, therefore, to maximize the sample sizes we only report the results for the samples without matching in $C$.

\begin{longrotatetable} 
\begin{deluxetable*}{crrccrcccrrccccc}
\tabletypesize{\footnotesize}
\tablecaption{Sample and Measurements \label{tab:result}} 
\tablewidth{11pc}
\tablehead{ 
\colhead{ Index } & 
\colhead{ R. A. } & 
\colhead{ Decl. } & 
\colhead{ $z$ } & 
\colhead{ log $M_*$ } &
\colhead{ log SFR } &
\colhead{ log $L_{\text{[O \textrm{\tiny III}]}}$ } &
\colhead{ Merger} &
\colhead{ $C$ } &
\colhead{ $A_{\rm CAS}$} &
\colhead{ $A_{\rm outer}$} &
\colhead{ $A_{\rm shape}$} &
\colhead{ $R_{\rm P}$} &
\colhead{ $R_{\rm max}$} &
\colhead{ $R_{\rm half}$} &
\colhead{ Type }
\\
\colhead{  } & 
\colhead{(\degree) } & 
\colhead{(\degree)  } & 
\colhead{  } & 
\colhead{ ($M_{\odot}$) } &
\colhead{ ($M_{\odot}$ yr$^{-1}$) } &
\colhead{ ($\mathrm{erg\,s^{-1}}$) } &
\colhead{ Flag } & 
\colhead{    } & 
\colhead{    } & 
\colhead{    } & 
\colhead{    } & 
\colhead{ (\arcsec) } & 
\colhead{ (\arcsec) } & 
\colhead{ (\arcsec) } &
\colhead{    } 
\\ 
\colhead{ (1) } &                                           
\colhead{ (2) } &                                            
\colhead{ (3) } &
\colhead{ (4) } &
\colhead{ (5) } &
\colhead{ (6) } &
\colhead{ (7) } &
\colhead{ (8) } &
\colhead{ (9) } &
\colhead{ (10) } &
\colhead{ (11) } &
\colhead{ (12) } &
\colhead{ (13) } &
\colhead{ (14) } &
\colhead{ (15) } &
\colhead{ (16) } 
}
\startdata
1 & 243.21553 & 0.37976 & 0.045 & 10.27 & 0.28 & \nodata  & 0 & 2.99 & 0.05 $\pm$ 0.07 & 0.01 $\pm$ 0.12 & 0.22 & 29.49 & 64.28 & 17.81 & SFG  \\
2 & 191.71165 & $-$0.73805 & 0.048 & 10.15 & 0.01 & \nodata  & 0 & 2.61 & $-$0.01 $\pm$ 0.04 & $-$0.02 $\pm$ 0.05 & 0.17 & 30.54 & 44.59 & 16.36 & SFG  \\
3 & 240.29665 & $-$0.82616 & 0.058 & 10.73 & 0.82 & \nodata  & 0 & 3.76 & 0.02 $\pm$ 0.01 & 0.03 $\pm$ 0.03 & 0.24 & 36.30 & 63.79 & 17.52 & SFG  \\
4 & 202.05917 & $-$0.39532 & 0.055 & 10.53 & 0.07 & \nodata  & 0 & 3.56 & $-$0.01 $\pm$ 0.07 & $-$0.04 $\pm$ 0.10 & 0.20 & 27.95 & 67.27 & 17.71 & SFG  \\
5 & 202.91333 & $-$0.33779 & 0.056 & 10.32 & $-$0.02 & \nodata  & 0 & 2.43 & 0.00 $\pm$ 0.01 & $-$0.02 $\pm$ 0.02 & 0.18 & 20.68 & 32.56 & 10.95 & SFG  \\
6 & 221.38405 & $-$1.15580 & 0.041 & 10.39 & $-$0.75 & 39.83 & 0 & 3.91 & 0.03 $\pm$ 0.02 & $-$0.01 $\pm$ 0.03 & 0.19 & 31.51 & 73.17 & 16.69 & AGN2  \\
7 & 231.17522 & $-$0.81717 & 0.052 & 10.53 & 0.19 & 39.99 & 0 & 2.97 & 0.01 $\pm$ 0.02 & 0.01 $\pm$ 0.05 & 0.25 & 21.51 & 55.89 & 14.94 & AGN2  \\
8 & 244.38397 & $-$0.28378 & 0.057 & 10.17 & 0.40 & 41.18 & 0 & 2.99 & 0.01 $\pm$ 0.01 & $-$0.01 $\pm$ 0.04 & 0.19 & 10.17 & 26.42 & 4.96 & AGN2  \\
9 & 240.82886 & 0.02428 & 0.041 & 10.17 & $-$0.07 & 39.83 & 0 & 3.39 & 0.01 $\pm$ 0.02 & $-$0.00 $\pm$ 0.04 & 0.19 & 21.76 & 58.49 & 13.67 & AGN2  \\
10 & 214.25699 & 0.47709 & 0.052 & 10.90 & 0.27 & 39.97 & 0 & 3.43 & 0.03 $\pm$ 0.02 & 0.01 $\pm$ 0.03 & 0.27 & 59.02 & 113.90 & 30.47 & AGN2  \\
11 & 10.65359 & $-$10.82274 & 0.042 & 10.34 & \nodata & 41.42 & 0 & 3.60 & 0.09 $\pm$ 0.01 & 0.05 $\pm$ 0.04 & 0.23 & 18.70 & 45.64 & 8.41 & AGN1  \\
12 & 12.85738 & $-$10.44575 & 0.055 & 10.23 & \nodata & 39.97 & 0 & 3.26 & 0.02 $\pm$ 0.06 & $-$0.01 $\pm$ 0.11 & 0.36 & 50.53 & 82.09 & 22.85 & AGN1  \\
13 & 14.69785 & $-$1.09716 & 0.047 & 10.47 & \nodata & 41.02 & 0 & 3.86 & $-$0.00 $\pm$ 0.06 & 0.00 $\pm$ 0.09 & 0.19 & 29.54 & 48.62 & 13.26 & AGN1  \\
14 & 20.49926 & $-$1.04010 & 0.054 & 10.70 & \nodata & 41.76 & 0 & 4.56 & 0.11 $\pm$ 0.02 & 0.08 $\pm$ 0.05 & 0.18 & 31.32 & 74.67 & 14.25 & AGN1  \\
15 & 120.68083 & 31.06759 & 0.041 & 10.88 & \nodata & 40.82 & 0 & 3.81 & 0.04 $\pm$ 0.03 & 0.06 $\pm$ 0.07 & 0.18 & 22.97 & 44.31 & 9.54 & AGN1  \\
\enddata
\tablecomments{ 
Column (1): Index number.
Column (2): Right ascension. 
Column (3): Declination. 
Column (4): Redshift.
Column (5): Stellar mass of the host galaxy.
Column (6): Star formation rate. 
Column (7): [O{\sevenrm\,III}] luminosity, corrected for Galactic but not internal extinction.
Column (8): Merger flag: 0 = no candidate major merger pair; 1 = at least one candidate major merger pair.
Column (9): Optical concentration.
Column (10): CAS asymmetry.
Column (11): Outer asymmetry.
Column (12): Shape asymmetry. 
Column (13): Petrosian radius.
Column (14): The semi-major axis of the minimal ellipse (with fixed center, elongation and orientation) that contains all of the main segment of the shape asymmetry segmentation map.
Column (15): The semi-major axis of an elliptical aperture that contains 50\% of the light.
Column (16): Subsample type. 
\\
(The full machine-readable table can be found in the online version.)
}
\end{deluxetable*}
\end{longrotatetable}

\subsection{Data Selection and Choice of Filter}

We download the images from the Pan-STARRS1 data archive\footnote{\url{https://ps1images.stsci.edu/cgi-bin/ps1cutouts}}, adopting the stacked images combined from multiple exposures \citep{Waters2020ApJS}, as they provide the best S/N.  We select image cutouts with a size that varies according to the dimension of each target galaxy, choosing 20\,\rpet\ to ensure that there is ample source-free region to measure the background accurately.  The Petrosian radius \rpet, which is insensitive to the depth of the image and the S/N of the source \citep{Rodriguez-Gomez2018MNRAS}, is obtained from the SDSS DR16 database\footnote{\url{https://www.sdss.org/dr16/data_access/}} instead of from the Pan-STARRS catalog, which lacks \rpet\ measurements for $\sim 10\%$ of our targets.  Comparison of the size measurements for the objects in common between the two catalogs reveals that \rpet\ from the SDSS catalog is on average $\sim 30\%$ lower than that measured from Pan-STARRS images.  We fit a linear relation between the two sets of data and use it to convert the SDSS-based values of \rpet\ to the Pan-STARRS zero point.

Our analysis is based on $i$-band images, whose red bandpass offers the best compromise in terms of sensitivity to stellar mass, minimization of the effects of dust absorption, and PSF quality, while at the same time avoiding the significant loss of sensitivity of the noisier $z$ and $y$ bands.  We explored the possibility of increasing the surface brightness sensitivity by stacking the images of all five Pan-STARRS bands, and various combinations thereof, but we finally abandoned the idea after discovering that the stacked images occasionally introduced artifacts due to the uneven background and ghost features, which can introduce spurious results in our asymmetry measurements.  For our purposes, the stacked images offered no advantage compared to the $i$-band images.

\begin{figure*}[htbp]
\centering
\includegraphics[width=0.98\textwidth]{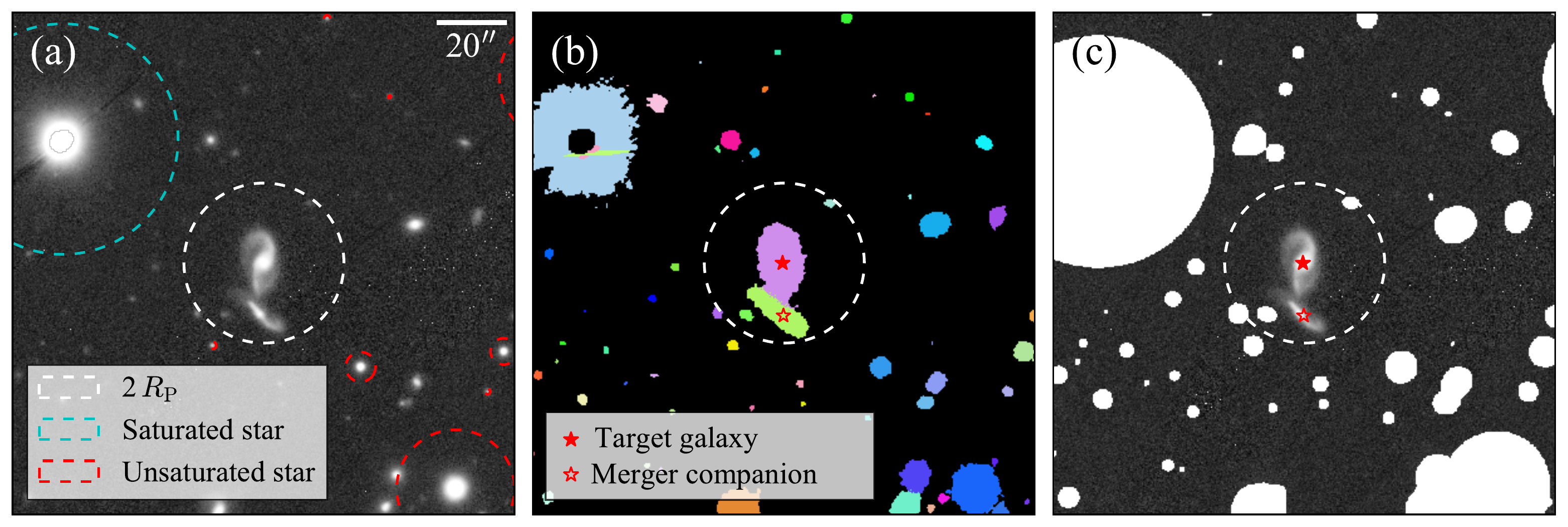}
\caption{(a) Example image of a target galaxy. The white dashed circle indicates $2\,\rpet$. Red and blue dashed circles mark the areas contaminated by unsaturated and saturated stars. Their masks are described by {\it mask 1}\ and {\it mask 2}, respectively.  (b) Segmentation map comprising {\it mask 3}\ outside $2\,\rpet$ and {\it mask 4}\ inside $2\,\rpet$.  (c) Example image after all of the contaminated pixels have been masked by our final mask.  Solid and open red stars mark the target galaxy and its candidate major merger companion, respectively, where the latter has a flux at least $1/4$ of that of the primary target.}
\label{Fig:Mask}
\end{figure*}

\begin{deluxetable*}{c | c | c | c | c}
\tablecaption{{\tt SExtractor} Parameters for Generating Masks} 
\label{tab:mask}
\tablehead{                                                                    
\colhead{}  & 
\multicolumn2c{Stars}  & 
\multicolumn2c{Galaxies} 
\\ 
\cline{2-5}
\colhead{Parameters} &   
\colhead{Unsaturated} & 
\colhead{Saturated} &                                 
\colhead{Outside $2\,\rpet$} &                                        
\colhead{Inside $2\,\rpet$}
\\
\colhead{} &                                      
\colhead{{\it mask 1}} &
\colhead{{\it mask 2}} &                                                  
\colhead{{\it mask 3}} &                                    
\colhead{{\it mask 4}}
}
\startdata
\texttt{DETECT\_THRESH}   & \multicolumn{4}{c}{$1\,\sigma$} \\ \hline
\texttt{DETECT\_MINAREA}  & \multicolumn{4}{c}{5}         \\ \hline
\texttt{DEBLEND\_NTHRESH} & \multicolumn{4}{c}{32}        \\ \hline
\texttt{BACK\_SIZE}       & 32    & 256   & 256   & 256   \\ \hline
\texttt{DEBLEND\_MINCONT} & 0.001 & 1 & 0.001 & 0.005     
\enddata
\end{deluxetable*}



\subsection{Generating Segmentation Map and Mask}
\label{ssec:mask}

As our galaxies are typically not very extended ($\sim 30-50$ pixels in diameter), determining the image background is not a critical step for the asymmetry measurements, especially since the background of Pan-STARRS images is usually relatively flat.  Nevertheless, we recalculate and subtract the median flux of the background, after iteratively clipping the source-free image beyond $3\,\sigma$, where $\sigma$ is the standard deviation of the background pixel distribution, using an initial mask generated with the function \texttt{make\_source\_mask} in the Python package \texttt{Photutils} \citep{Bradley2020}.  We do not attempt higher-order fitting of the background in order to avoid oversubtraction of real signal from the extended outskirts of galaxies.  

For the actual measurement of asymmetry, we need a more elaborate procedure to generate a mask that not only carefully flags contaminating sources and image artifacts but also retains, if present, probable physical companions and extended features arising from ongoing mergers and tidal interactions \citep{Toomre&Toomre1972,Barnes1992ApJ,Struck2012,Duc2013}.  For the majority of the non-stellar sources in the field, we do not know whether they are associated with the primary target because they usually lack spectroscopic redshifts, and photometric redshifts are both incomplete and not accurate enough. Here we assume that any extended source projected within 2~\rpet\ of the primary target having at least 25\% of its flux to be a candidate companion galaxy belonging to a major merger, which is normally defined by a mass ratio of 1:4 \citep{Bournaud2005aap, Conselice2014ARA&A}.  We make the simplifying assumption that the flux ratio in the $i$ band adequately approximates the stellar mass ratio.

We use \texttt{SExtractor} \citep{Bertin1996A&AS}\footnote{We verified that \texttt{Photutils} is equally effective.} to generate an image segmentation map to detect and deblend sources. This code can generate a segmentation map with different levels of complexity depending on the properties of the source and the goal of the detection.  This flexibility is necessary to cope with the many faint, small sources that are often projected close to and overlap with the brighter, more extended target of interest.  Contamination by foreground stars, whether saturated or not, also poses a challenge.  Following common practice (e.g., \citealt{Gray2009MNRAS,Galametz2013ApJS,Sazonova2021ApJ}), we use the ``cold'' mode to identify the pixels of bright, extended sources using a high detection threshold and a large minimum connecting area with proper deblending.  Faint, small sources are not detected in this mode.  Meanwhile, a ``hot'' mode that uses a low detection threshold and a small minimum connecting area provides complementary information, at the expense of sometimes breaking bright sources (e.g., a highly structured spiral galaxy) into many pieces \citep{,Barden2012MNRAS}.  The final source catalog then comprises objects detected in the cold mode plus sources found in the hot mode where no cold source was detected.  The mask can then be generated by combining all the source segments in the final catalog.  We develop a new method that can detect all sources comprehensively and generate a mask based on the area influenced by them.  We combine segmentation maps generated by \texttt{SExtractor} using four modes, each
designed to identify and mask a specific kind of contaminant.

\texttt{SExtractor} considers a source detected if there are at least \texttt{DETECT\_MINAREA} connected pixels above the detection threshold \texttt{DETECT\_THRESH}, which is in units of the background noise ($\sigma$).  A low detection threshold ($\texttt{DETECT\_THRESH}=1$) is effective to find all faint sources as well as to capture the outskirts of bright stars.  A source is classified as a star or a galaxy based on a neural network algorithm \citep{Nonino1999SE}, with the likelihood of it being a star indicated by the parameter \texttt{CLASS\_STAR} approaching unity.  In common for all source types, we set $\texttt{DETECT\_THRESH}=1$ and $\texttt{DETECT\_MINAREA}=5$.  Two parameters are important for deblending sources in a crowded field: the deblending threshold ($\texttt{DEBLEND\_NTHRESH}$) controls the level of deblending, and, following the suggestion of the \texttt{SExtractor} manual, we set it to 32 to achieve strong deblending; $\texttt{DEBLEND\_MINCONT}$, the flux contrast to deblend several overlapping objects, is set differently for different types of sources (Table~\ref{tab:mask}).  We set the background mesh size to $\texttt{BACK\_SIZE}=256$ pixels when we want to detect the outskirts of a galaxy, and to 32 pixels for point sources, both within the range suggested by \cite{Bertin1996A&AS}.  

All the sources should be masked except for the target galaxy and, if present, its possible major merger companion, which is defined as a non-stellar source within $2\,\rpet$ from the central target having an isophotal flux that exceeds 25\% of the flux of the main target.  Figure~\ref{Fig:Mask} gives an example image of a target galaxy, its segmentation map, and final mask.  We classify the detected sources into four categories, each of which is treated with a separate type of mask: 

\begin{enumerate}
\item \textit{Unsaturated stars (mask 1)}\label{segm1}.
We first generate the segmentation map with $\texttt{BACK\_SIZE}=32$ and deblend with contrast $\texttt{DEBLEND\_MINCONT} = 0.001$. This step provides the list of all detectable sources in the image.  We mask the point sources (unsaturated stars) with $\texttt{CLASS\_STAR}>0.75$.  The extended sources, which may be split up by the aggressive deblending, are treated in subsequent steps.  To fully mask the wings of the PSF, we model the one-dimensional azimuthally averaged surface brightness profile of the Pan-STARRS $i$-band PSF with a \cite{Moffat1969A&A} function and estimate the full extent of the region influenced by the star.

\item \textit{Saturated stars (mask 2)}\label{segm2}. 
Saturated stars are characterized by a central core flattened by the saturation limit and highly extended wings from the PSF.  We choose $\texttt{DEBLEND\_MINCONT}=1$ to preclude deblending, so as to guarantee the integrity of the segment of the saturated star, and a large background mesh size of $\texttt{BACK\_SIZE}=256$ in order to enclose the outskirts of the extended emission.  Instead of generating the mask based on the isophotal flux, which is underestimated for saturated stars, we use the isophotal segmentation area (\texttt{ISOAREA\_IMAGE}) to trace the region influenced by the saturated stars, adopting an empirical scaling between this area and the radius of the circular mask.

\item \textit{Galaxies at $>2 \,\rpet$ (mask 3)\label{segm3}}. To mask completely the galaxies outside $2 \,\rpet$, we again adopt $\texttt{BACK\_SIZE}=256$ and deblend the segmentation map with $\texttt{DEBLEND\_MINCONT} = 0.001$ so as to separate the target from neighboring galaxies.  We then grow the segmentation map outside $2 \,\rpet$ by convolving it with a Gaussian kernel with standard deviation 3 times the PSF FWHM (Figure \ref{Fig:Mask}c). 

\item \textit{Galaxies at $<2 \,\rpet$ (mask 4)\label{segm4}}.  Within $2\,\rpet$, we need to mask all stars and faint galaxies, but keep the target and its possible major merger companion, if present.  The faint galaxies can be well masked by an elliptical aperture with semi-major and semi-minor axis equal to 3 times those measured by \texttt{SExtractor}.  To prevent over-blending of fine structures 
such as tidal tails and spiral arms, we adopt $\texttt{DEBLEND\_MINCONT} = 0.005$, which has moderate deblending effect, and maintain \texttt{BACK\_SIZE} = 256. 
\end{enumerate}

\begin{figure*}
\centering
\includegraphics[width=0.98\textwidth]{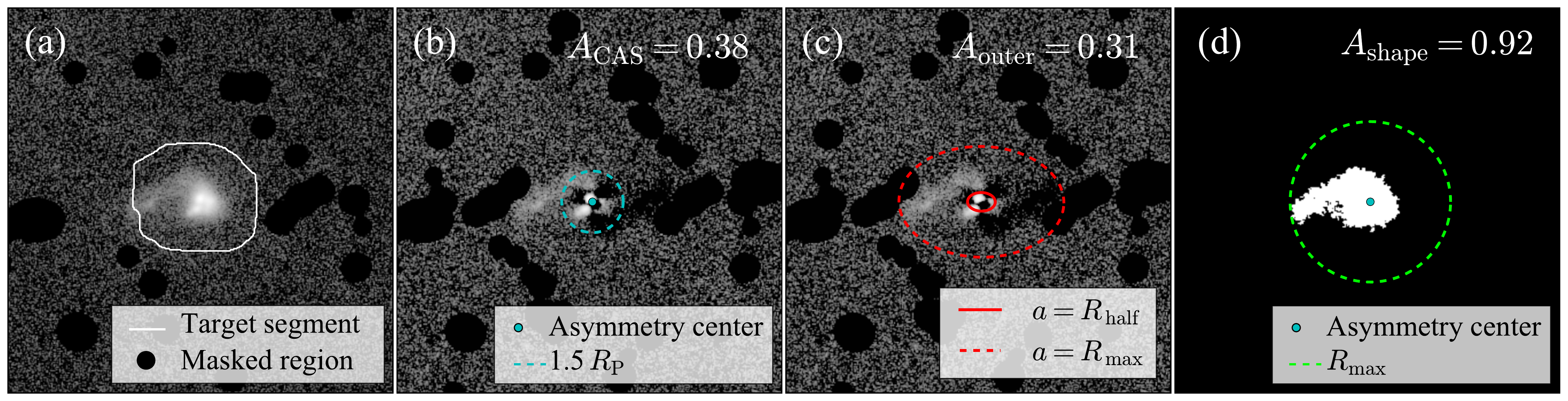}
\caption{Analysis of an example star-forming galaxy using \statm. (a) Example image after the contaminated pixels have been masked by our final mask. The target galaxy segmentation map is indicated by white contour. (b) Residual image by subtracting the rotated image from the original image, $I- I_{\mathrm{180}}$, around the asymmetry center marked by the blue dot.  \acas\ is integrated inside a circular region with radius $1.5\,\rpet$ (blue dash aperture). (c) Same background as panel (b), overlaid by an ellipse with semi-major axis $a=R_{\rm half}$ (solid red) and another with $a=R_{\rm max}$ (dashed red);  \aout\ is integrated between the two ellipses.  (d) Segmentation map for calculating \asha\ within the green circle with radius $R_{\rm max}$.}
\label{fig:stat_result}
\end{figure*}

\section{Asymmetry Measurements} 
\label{sec4:asy}    

We use the Python package \statm\ \citep{Rodriguez-Gomez2018MNRAS}, with some modifications, to calculate the asymmetry parameters.  The code requires a background-subtracted image, a mask image to identify the sources to be ignored, and the segmentation map, which gives the position of the target and is used to compute source properties such as its ellipticity and position angle.  The segmentation map derives from \textit{mask 4}, after smoothing it with a box kernel with size equal to $2\,\rpet$ of the target galaxy to regularize it \citep{Sazonova2021ApJ}.  Three types of asymmetry parameters can be calculated by \statm, and we use all of them in our analysis.  

The ``CAS'' asymmetry is calculated by subtracting the target image with a 180\degree-rotated image of itself \citep{Schade1995ApJ,Abraham1996ApJS,Conselice2000ApJ}:
    
\begin{equation} \label{eq:Asymmetry}
A=\frac { { \Sigma  }_{ i,j }|{ I }_{ ij }-{ I }_{ ij }^{ 180 }| }{ { \Sigma  }_{ i,j }|{ I }_{ ij }| } - \frac{N_\mathrm{src}\,\abkg}{{ \Sigma  }_{ i,j }|{ I }_{ ij }|},
\end{equation}
    
\noindent
where ${ I }_{ ij }$ and ${ I }_{ ij }^{ 180 }$ are the pixel flux value and the one rotated by 180\degree, respectively, ${ A }_\mathrm{ bgk }$ is the pixel-wise averaged asymmetry of the background around the source, and $N_\mathrm{src}$ is the number of unmasked pixels of the image of the source.  The CAS asymmetry is calculated within $1.5 \,\rpet$ (Figure~\ref{fig:stat_result}b). The rotation 
center is the position that minimizes the asymmetry value.  Instead of measuring ${ A }_\mathrm{ bgk }$ within a single ``sky box'' that \statm\ adopts, we calculate it by randomly sampling the background 100 times, requiring that there be $<10\%$ masked pixels inside the sample box (see Appendix~\ref{apd:unc} for more details).  We adopt a sample box size comparable to the size of the target.

The outer asymmetry \citep{Wen2014ApJ}, \aout, also can be defined by Equation~(\ref{eq:Asymmetry}).  Instead of summing the pixels over the entire region within 1.5\,\rpet, we only sum the pixels inside an elliptical annulus in the outskirts of the galaxy.  The inner and outer semi-major axes of the elliptical annulus are $R_{\rm half}$ and $R_{\rm max}$ (the red solid and dashed ellipses in Figure \ref{fig:stat_result}c), where $R_{\rm max}$ is the maximum radius of the target's emission, which is defined rigorously below for the shape asymmetry, and $R_{\rm half}$ is the radius that contains half of the total flux within $R_{\rm max}$.  The ellipticity and position angle of the annulus are determined by the input segmentation map of the target \citep{Rodriguez-Gomez2018MNRAS}.  The rotation center and ${ A }_\mathrm{ bgk }$ are identical to those used for the CAS asymmetry.

In order to improve the sensitivity to the low surface brightness features of the galaxy, the shape asymmetry (\asha; \citealt{Pawlik2016MNRAS}) is calculated using the same mathematical form as Equation~(\ref{eq:Asymmetry}), but based on the ``binary segmentation map'' (Figure~\ref{fig:stat_result}d), where the values of pixels belonging to the target galaxy are 1, and the values of background pixels are 0.  The key to calculating the shape asymmetry calculation is separating target pixels from background pixels.  First, the background noise is estimated within a circular annulus between 2 and 4 times \rpet. Within the annulus, pixels with flux value above $3\,\sigma$ are iteratively clipped, until the \texttt{mode} value converges to $\texttt{mode}=2.5 \times \texttt{median}  - 1.5 \times \texttt{mean}$ \citep{Bertin1996A&AS}. Second, contiguous pixels above the \texttt{mode} by $1 \,\sigma$ are considered as the target galaxy pixels, and they are marked with value 1; the rest are marked with value 0.  Lastly, we smooth this map by a $3 \times 3$ boxcar filter to obtain the final binary segmentation map.  The rotation center is the same as that determined for the CAS asymmetry; the background asymmetry ($A_\mathrm{bgk}$) is, by definition, 0.  We calculate \asha\ within a radius $R_{\rm max}$ (green circle in Figure~\ref{fig:stat_result}d), the distance of the furthest galaxy pixel from the center.

During the course of making the asymmetry measurements, we flag and remove from further consideration targets that fall under any of the following conditions: (1) the target center is masked; (2) more than 20\% of the pixels within any of the regions used to calculate the three asymmetry parameters are masked; (3) more than 20\% of the pixels inside the input target segment region are masked; and (4) the asymmetry center is incorrectly determined and far from the target.  A total of 11\%, 12\%, and 10\% of the sources were removed for these reasons from the SFG, AGN2, and AGN1 samples, respectively. We verified through the K-S test that the remaining samples are still well-matched in redshift and stellar mass.

\begin{figure*}
\centering
\includegraphics[width=0.98\textwidth]{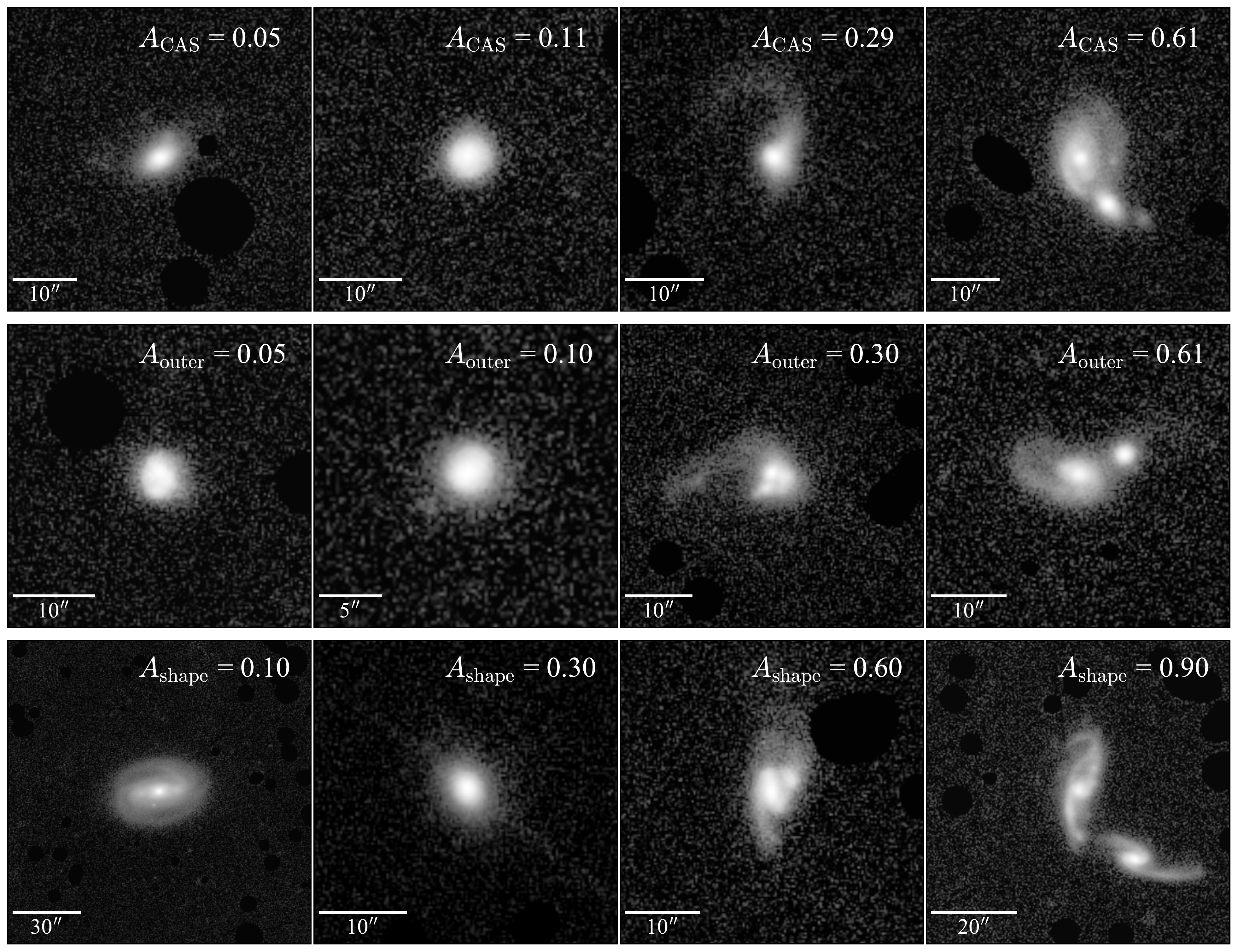}
\caption{Example star-forming galaxies, showing (top) \acas, (middle) \aout, and (bottom) \asha, with asymmetry values increasing from left to right, as labeled on the upper-right corner of each panel.}
\label{fig:img_sf}
\end{figure*}

\begin{figure*}
\centering
\includegraphics[width=0.98\textwidth]{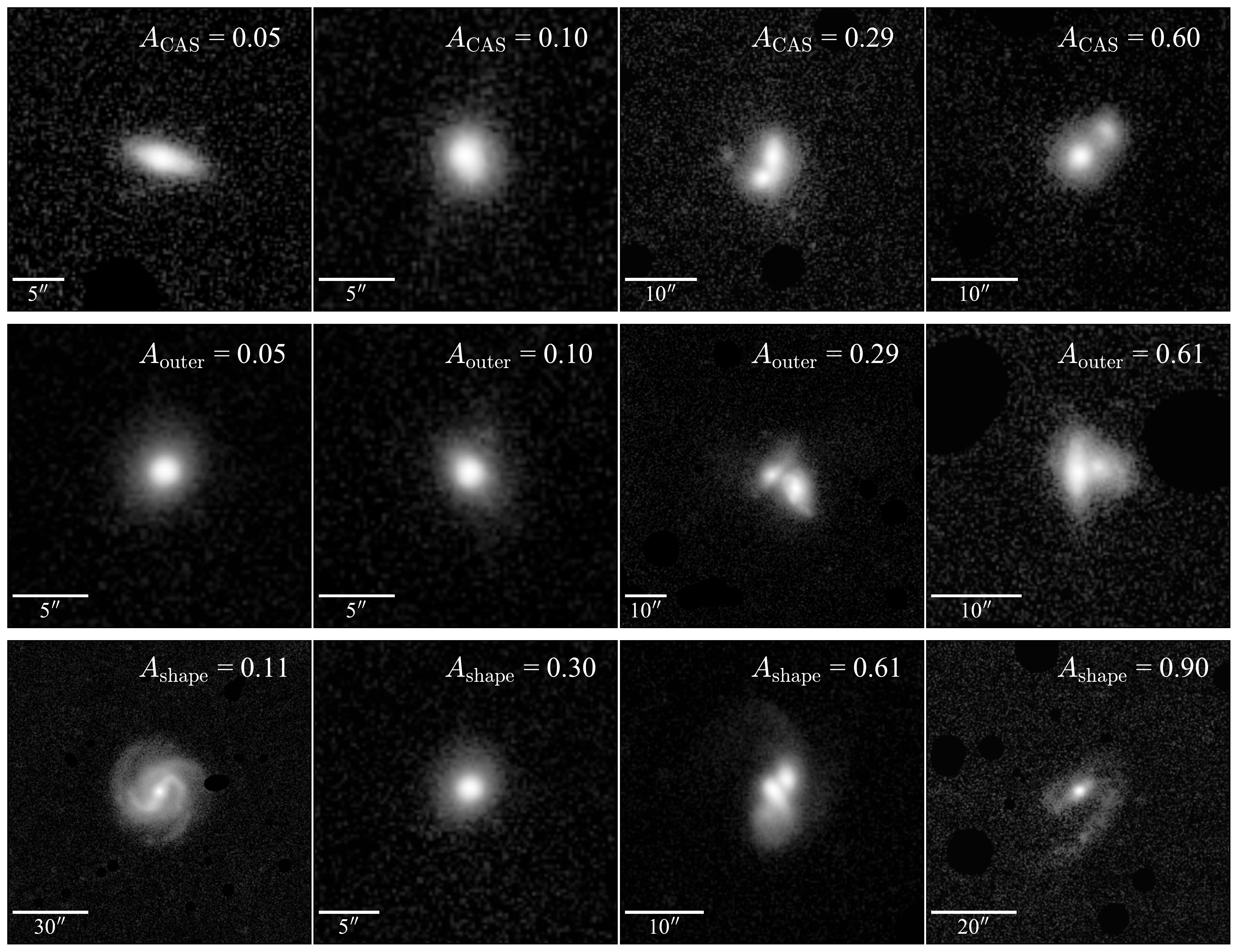}
\caption{Example type~2 AGNs, showing (top) \acas, (middle) \aout, and (bottom) \asha, with asymmetry values increasing from left to right, as labeled on the upper-right corner of each panel.}
\label{fig:img_AGN2}
\end{figure*}

\begin{figure*}
\centering
\includegraphics[width=0.98\textwidth]{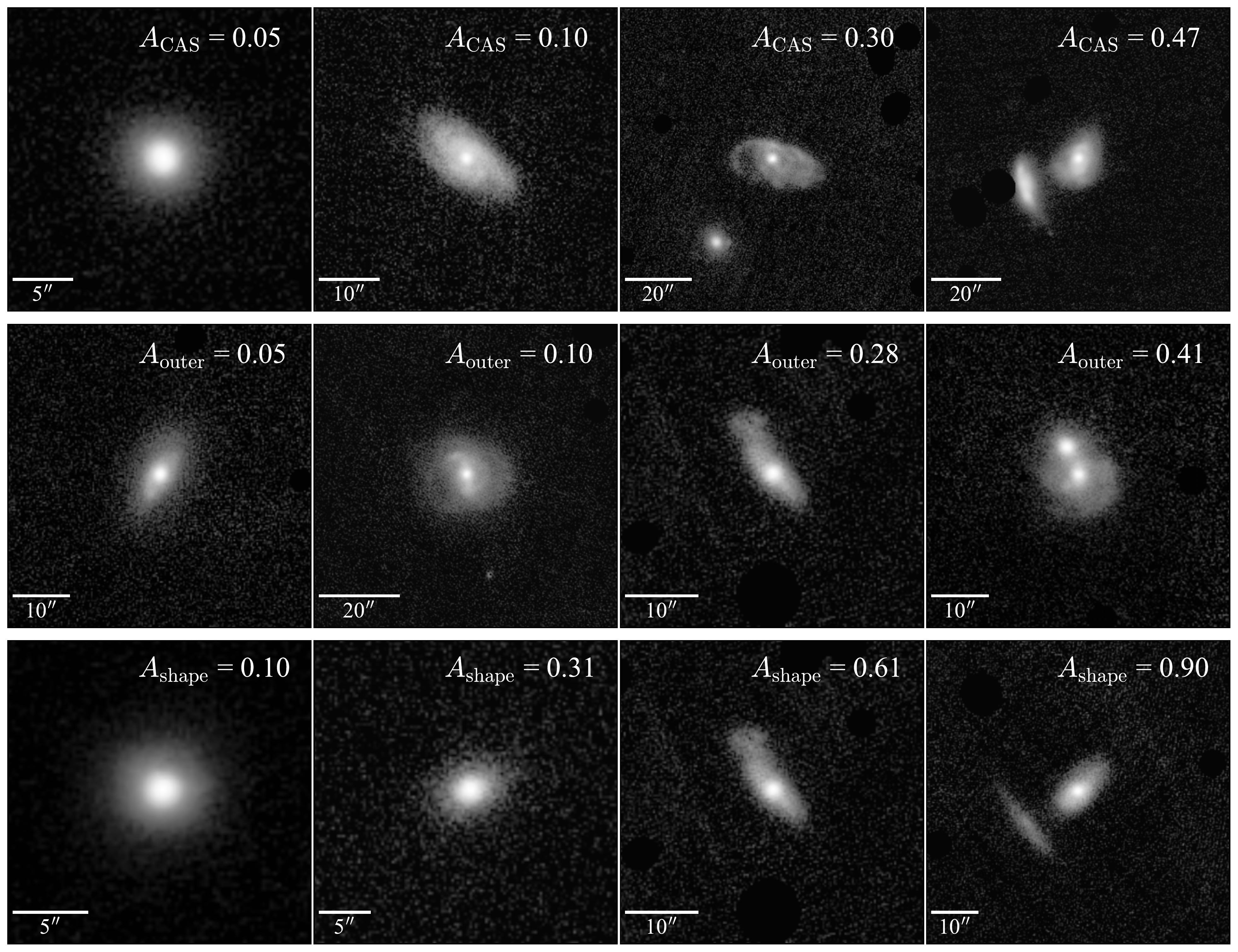}
\caption{Example type~1 AGNs, showing (top) \acas, (middle) \aout, and (bottom) \asha, with asymmetry values increasing from left to right, as labeled on the upper-right corner of each panel.}
\label{fig:img_t1}
\end{figure*}

Figures~\ref{fig:img_sf}--\ref{fig:img_t1} give examples of the asymmetry parameters measured for the three galaxy samples.  The morphology of the galaxies becomes increasingly asymmetric as \acas, \aout, and \asha\ progress from low to high values.  The rightmost column of each figure, which showcases galaxies with the highest asymmetry, are mostly major mergers. The AGN1 sample shows clear pointlike nuclei in some targets, although they are typically not overwhelming.  This is qualitatively consistent with the overall consistent distribution of concentration parameters for the two AGN types (Figure~\ref{Fig:C_match}a).  As discussed in Appendix~\ref{apd:unc}, while the \acas\ parameter of AGN1s may be affected by nuclear emission, \aout\ should be robust.

\begin{figure*}[htbp]
\centering
\includegraphics[width=0.9\textwidth]{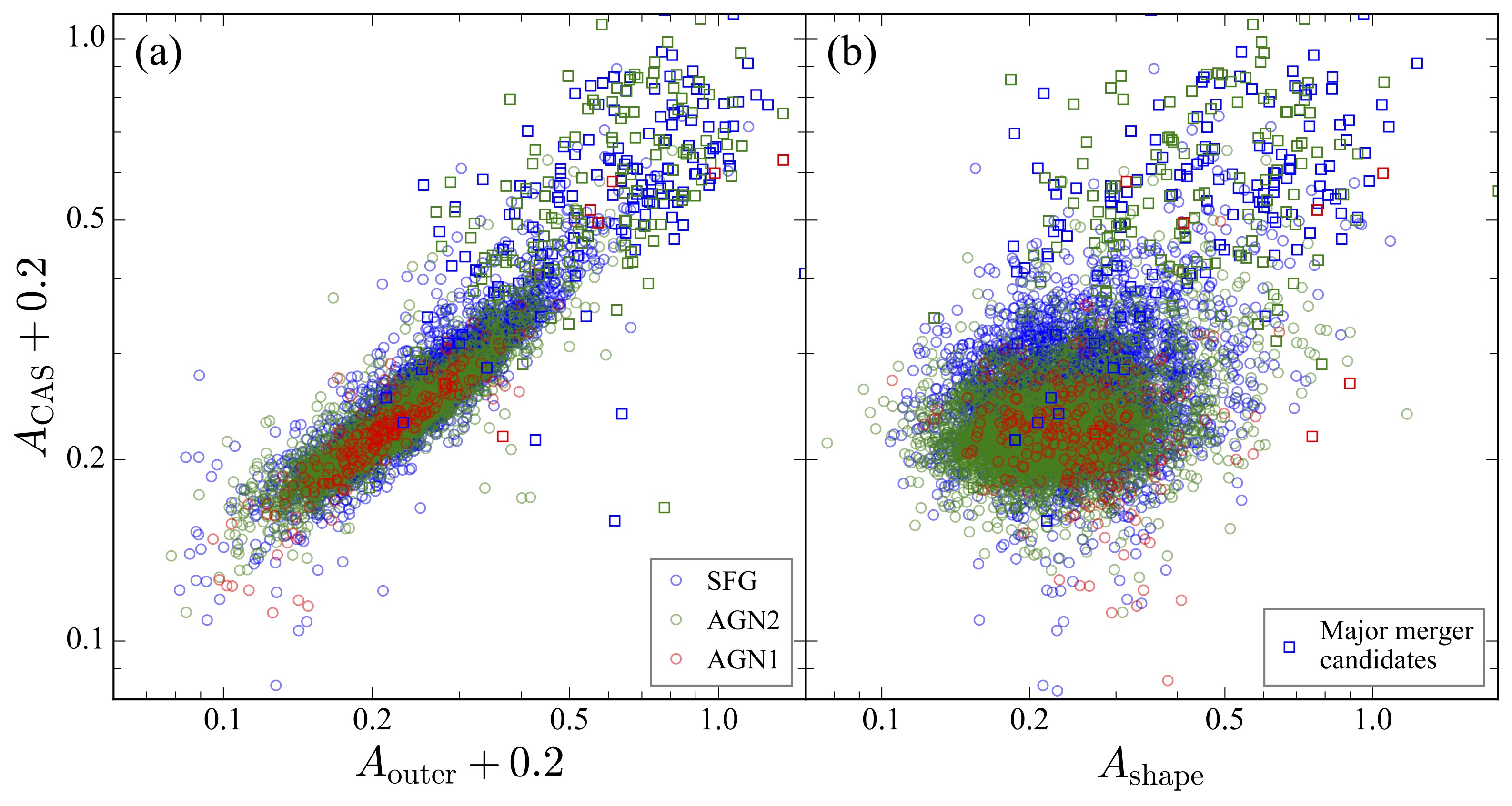}
\caption{\acas\ versus (a) \aout\ and (b) \asha\ for star-forming galaxies (SFG; blue), type~2 AGNs (AGN2; green), and type~1 AGNs (AGN1; red).  Major merger candidates are highlighted as squares.} 
\label{Fig:A_correlation}
\end{figure*}

Comparing the three asymmetry parameters (Figure~\ref{Fig:A_correlation}), we find very strong correlations between \acas\ and \aout\ for all three samples (Spearman correlation coefficient $\rho>0.9$ and $p<0.001$).  However, the correlations between \acas\ and \asha\ are quite weak for SFGs and AGN2s ($\rho \approx 0.2$, $p<0.001$), and it is insignificant for AGN1s ($\rho= 0.07$, $p=0.15$)\footnote{Throughout the paper, we consider a result to be statistically significant if the $p$-value is smaller than 0.01.  We prefer 0.01 to 0.05 as the threshold because our sample sizes are relatively large.}.  The parameter \asha, on account of its strong sensitivity to the noise of the background, is only useful to identify highly asymmetric sources and appears to be very uncertain when $\asha<0.5$, which characterizes most of our targets.  Therefore, in the rest of the paper we mainly focus on \acas\ and \aout\ for SFGs and AGN2s, and for AGN1s we only use \aout\ to avoid possible systematic bias on \acas\ due to the central nucleus (Appendix~\ref{apd:unc}).

\section{Results and Discussion} \label{sec5:res}

\subsection{Comparing the Asymmetries of Star-forming Galaxies and AGNs}
\label{ssec:comp}

\begin{figure*}[htbp]
\centering
\includegraphics[width=\textwidth]{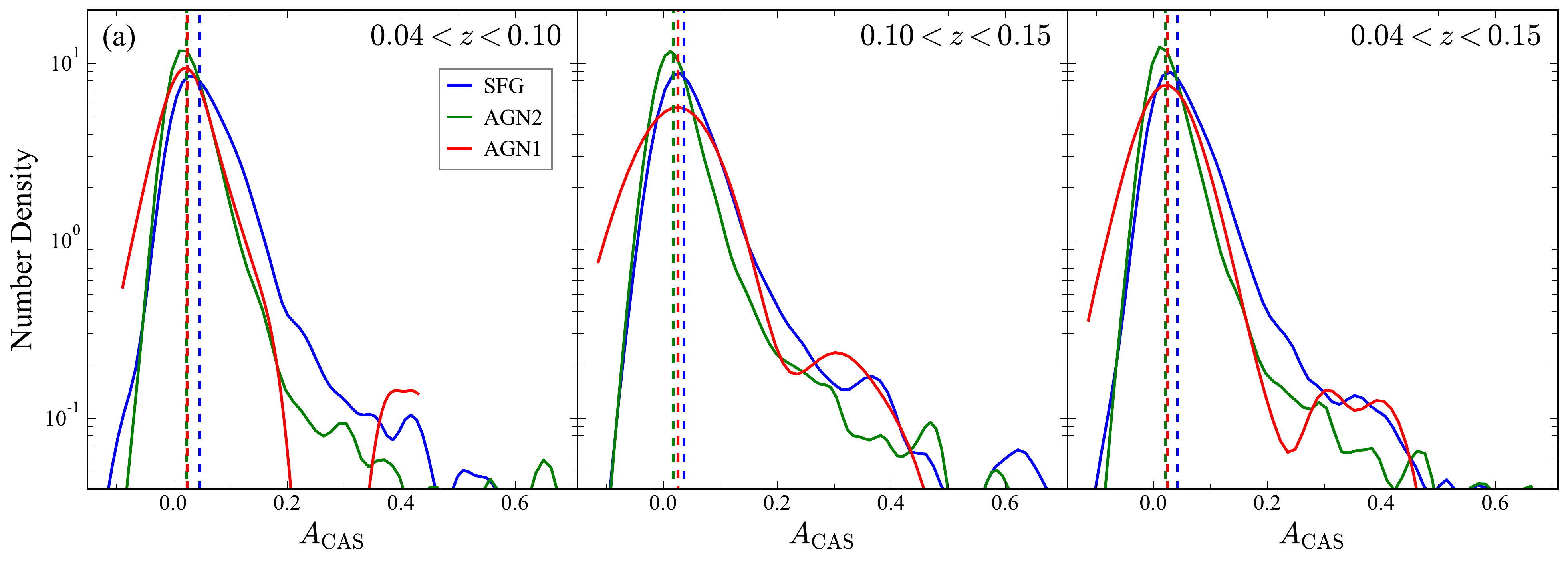}
\includegraphics[width=\textwidth]{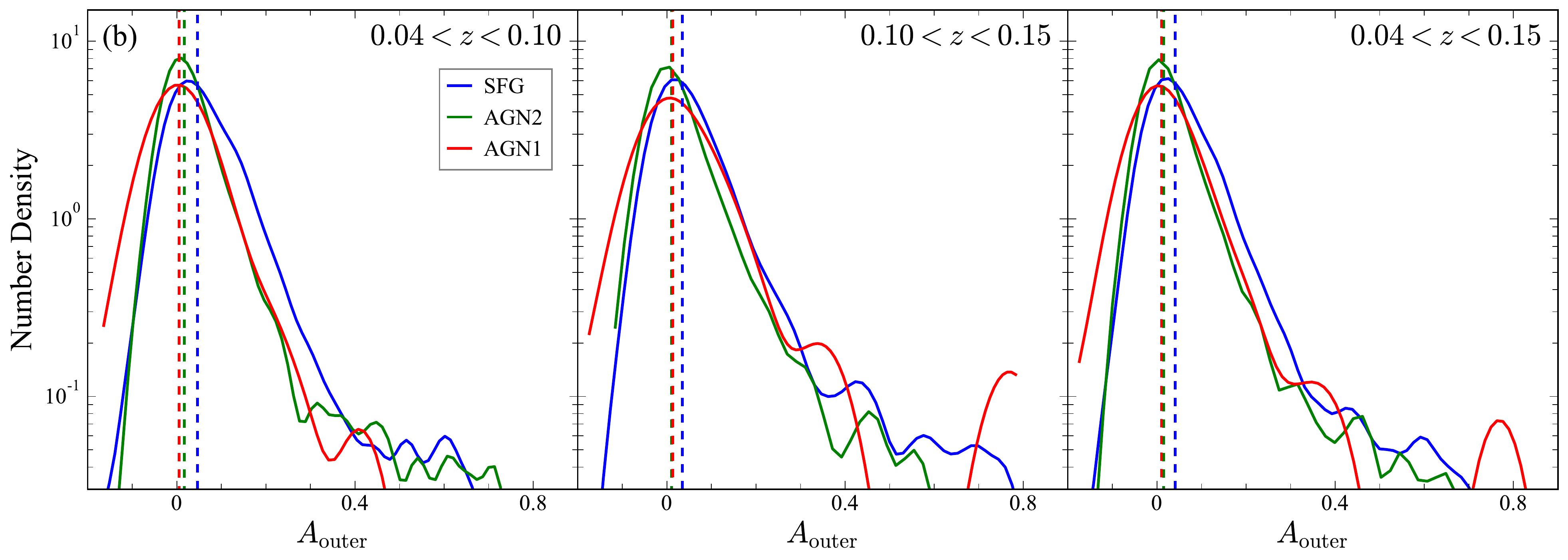}
\includegraphics[width=\textwidth]{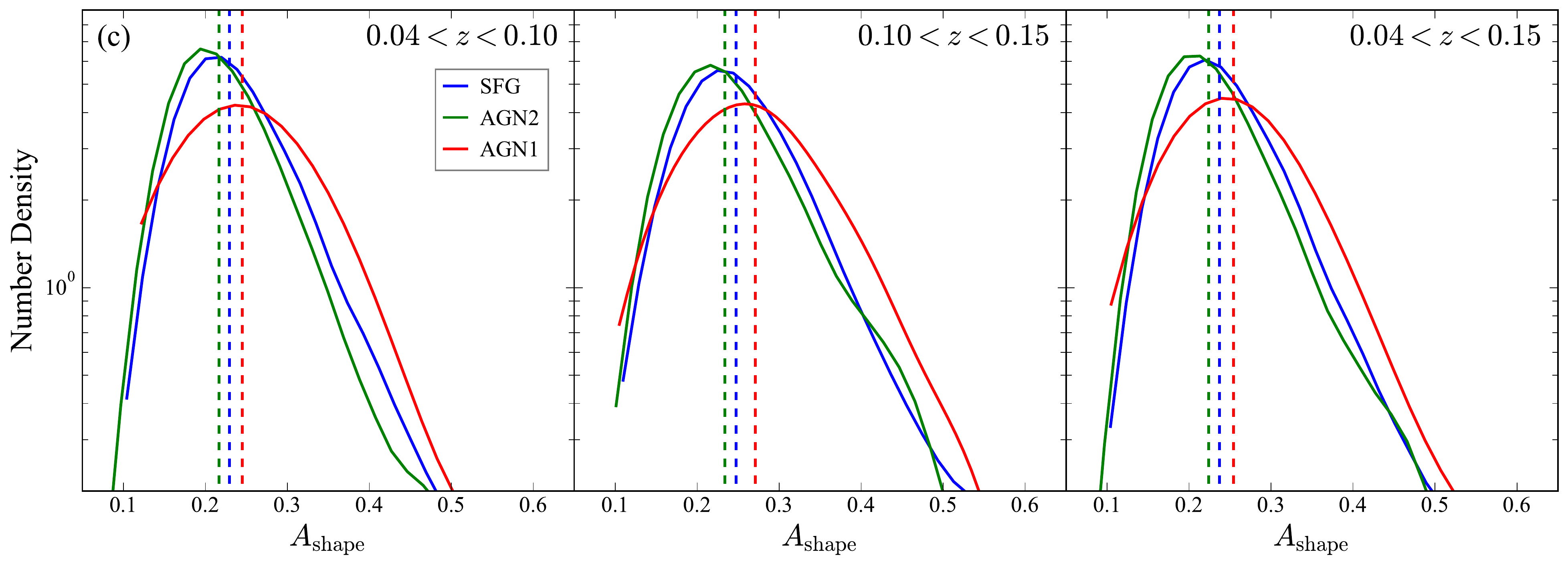}
\caption{Kernel-density estimates of (a) $A_{\rm CAS}$, (b) $A_{\rm outer}$, and (c) $A_{\rm shape}$ for star-forming galaxies (SFG; blue), type~2 AGNs (AGN2; green), and type~1 AGNs (AGN1; red), with median values indicated by the vertical dashed lines, for objects with (left) $0.04 < z < 0.10$, (middle) $0.10 < z < 0.15$, and (right) $0.04 < z < 0.15$ (whole sample).}
\label{Fig:hist}
\end{figure*}

Figure~\ref{Fig:hist} compares the measured asymmetry parameters of the three galaxy samples.  Somewhat to our surprise, the AGNs, whether both types combined or each type separately, actually show {\it lower}\ levels of asymmetry compared to SFGs, in all three asymmetry parameters, both for the whole sample covering all redshifts ($0.04<z<0.15$) and separately for the lower ($0.04<z<0.10$) and higher ($0.10<z<0.15$) redshift intervals.  According to the K-S test and Student's $t$-test (Table~\ref{tab:test_ks_t}), the null hypothesis that the asymmetry parameters of the AGN and SFG samples are drawn from the same parent population can be rejected with a probability of $p<10^{-3}$.  This can be seen in the slight but systematically lower mean and median values of the asymmetry in AGNs compared to SFGs (Table~\ref{tab:stat}).  AGNs also have somewhat narrower asymmetry distributions, as reflected in the lower standard deviations.  As mentioned in Section~\ref{sec2:data} (see Figure~\ref{Fig:C_match}), the minor differences in concentration between the AGNs and SFGs may be a potential source of concern when comparing active and inactive galaxies.  We have verified, but for the sake of brevity do not show, that none of our main results are changed using subsamples that are matched in $C$.

Our conclusions differ from those of \cite{Maiolino1997ApJ}, who found that Seyfert~2 galaxies are more likely to be interacting with a companion than are the field galaxies and Seyfert~1 galaxies.  The results of \cite{Maiolino1997ApJ} were based on visual inspection of a relatively small sample, which, in addition to being subjective, only pick up galaxy pairs or pre-mergers. Our objective non-parametric method is more sensitive to weak morphological distortions, and we have applied it to a large, homogeneous sample of AGNs and a carefully matched sample of control galaxies.  As discussed in Section~\ref{ssec:fmerg}, AGN2s do not exhibit a higher incidence of candidate major merger pairs than SFGs or AGN1s.  The results are not sensitive to redshift.  The typical half-light radii of our sources is $\sim 2.5$ times the PSF size, and we have verified that no strong dependence on redshift can be discerned (Figure~\ref{Fig:hist}), consistent with the tests performed by \cite{Thorp2021MNRAS} to evaluate the effect of resolution on asymmetry measurements.

Another key finding is that both AGN types have statistically similar degrees of asymmetry.  We base this conclusion only on the results derived from \aout, which, as argued in Section~\ref{sec4:asy} and Appendix A, provides the most reliable measure of asymmetry for AGN1s.  Focusing on the entire ($0.04<z<0.15$) sample, the null hypothesis that AGN1s and AGN2s derive from the same parent population can be rejected with a probability of $p=0.145$ according to the K-S test and $p=0.545$ according to Student's $t$-test (Table~\ref{tab:test_ks_t}).

\begin{deluxetable*}{rccccccccc}
\tabletypesize{\small}       
\tablecaption{Results of Kolmogorov-Smirnov Test and Student's $t$-test \label{tab:test_ks_t}}
\tablewidth{0pc} 
\tablehead{                                                                    
\colhead{}  &     
\colhead{}  & 
\multicolumn{2}{c}{  AGN1+AGN2/SFG} &                                                           
\colhead{ }  &                  
\multicolumn{2}{c}{  AGN2/SFG} &
\colhead{ }  &                               
\multicolumn{2}{c}{  AGN1/AGN2} 
\\ 
\cline{3-4} \cline{6-7} \cline{9-10}
\colhead{ $z$ } &   
\colhead{ Parameter} & 
\colhead{  K-S} &                                 
\colhead{  $t$} &                                        
\colhead{ } &                                                   
\colhead{  K-S} &                                      
\colhead{  $t$} &
\colhead{ } &                                                  
\colhead{  K-S} &                                    
\colhead{  $t$}
\\ 
\colhead{(1)} &                                           
\colhead{  (2)} &                                                     
\colhead{  (3) } & 
\colhead{  (4) } & 
\colhead{  } &   
\colhead{  (5) } & 
\colhead{  (6) } & 
\colhead{  } &                                                             
\colhead{  (7) } &
\colhead{  (8) }
}
\startdata
\multirow{3}{*}{0.04--0.10} & \acas     & $<10^{-3}$ & $<10^{-3}$ &  & $<10^{-3}$ & $<10^{-3}$ &  & 0.060      & 0.114     \\
                            & \aout     & $<10^{-3}$ & $<10^{-3}$ &  & $<10^{-3}$ & $<10^{-3}$ &  & 0.025      & 0.099     \\
                            & \asha     & $<10^{-3}$ & $<10^{-3}$ &  & $<10^{-3}$ & $<10^{-3}$ &  & $<10^{-3}$ & $<10^{-3}$ \\  \hline
\multirow{3}{*}{0.10--0.15} & \acas     & $<10^{-3}$ & $<10^{-3}$ &  & $<10^{-3}$ & $<10^{-3}$ &  & 0.046      & 0.300      \\
                            & \aout     & $<10^{-3}$ & $<10^{-3}$ &  & $<10^{-3}$ & $<10^{-3}$ &  & 0.516      & 0.439      \\
                            & \asha     & $<10^{-3}$ & 0.002      &  & $<10^{-3}$ & $<10^{-3}$ &  & $<10^{-3}$ & 0.018      \\  \hline
\multirow{3}{*}{0.04--0.15} & \acas     & $<10^{-3}$ & $<10^{-3}$ &  & $<10^{-3}$ & $<10^{-3}$ &  & 0.022      & 0.632      \\
                            & \aout     & $<10^{-3}$ & $<10^{-3}$ &  & $<10^{-3}$ & $<10^{-3}$ &  & 0.145      & 0.545      \\
                            & \asha     & $<10^{-3}$ & $<10^{-3}$ &  & $<10^{-3}$ & $<10^{-3}$ &  & $<10^{-3}$ & $<10^{-3}$
\enddata
\tablecomments{The $p$-values of the Kolmogorov-Smirnov test and Student's 
$t$-test on \acas, \aout, and \asha\ measurements of the SFG, AGN2, and AGN1 samples.  Our analysis 
mainly focuses on the \acas\ and \aout\ measurements of SFGs and AGN2s and the \aout\ results of
AGN1s because the other parameters are not robust (see Section~\ref{sec4:asy} 
and Appendix~\ref{apd:unc}).
Column (1): Redshift range. 
Column (2): The three asymmetry parameters. 
Columns (3)--(4): Comparisons between the AGN1+AGN2 and SFG samples.
Columns (5)--(6): Comparisons between the AGN2 and SFG samples.
Columns (7)--(8): Comparisons between the AGN1 and AGN2 samples.
}
\end{deluxetable*}

\begin{deluxetable*}{c c c c c c c c c c c c c c c c c}
\tablecaption{Statistics of the Asymmetry Parameters} 
\label{tab:stat}
\tablehead{                                                                    
\colhead{\multirow{2}{*}{$z$}} & 
\colhead{\multirow{2}{*}{Parameter}} & 
\multicolumn3c{Number} &
\colhead{} &
\multicolumn3c{Mean} &
\colhead{} &
\multicolumn3c{$\sigma$} &
\colhead{} &
\multicolumn3c{Median}
\\ 
\cline{3-5} \cline{7-9} \cline{11-13}  \cline{15-17}
\colhead{} &   
\colhead{} & 
\colhead{SFG} &                                 
\colhead{AGN2} &                                        
\colhead{AGN1} &
\colhead{} &                                 
\colhead{SFG} &                                 
\colhead{AGN2} &                                        
\colhead{AGN1} &
\colhead{} &                                 
\colhead{SFG} &                                 
\colhead{AGN2} &                                        
\colhead{AGN1} &
\colhead{} &                                 
\colhead{SFG} &                                 
\colhead{AGN2} &                                        
\colhead{AGN1} \\
\colhead{(1)} &   
\colhead{(2)} & 
\colhead{(3)} &                                 
\colhead{(4)} &                                        
\colhead{(5)} &
\colhead{} &                                 
\colhead{(6)} &                                 
\colhead{(7)} &                                        
\colhead{(8)} &
\colhead{} &                                 
\colhead{(9)} &                                 
\colhead{(10)} &                                        
\colhead{(11)} &
\colhead{} &                                 
\colhead{(12)} &                                 
\colhead{(13)} &                                        
\colhead{(14)} 
}
\startdata
\multirow{3}{*}{0.04--0.10} & \acas      & \multirow{3}{*}{2542} & \multirow{3}{*}{2536} & \multirow{3}{*}{136} & & 0.07 & 0.04 & 0.03 && 0.09 & 0.09 & 0.08 && 0.05 & 0.03 & 0.03 \\
                            & \aout      &                       &                       &                      & & 0.07 & 0.04 & 0.02 && 0.13 & 0.12 & 0.15 && 0.05 & 0.02 & 0.01 \\
                            & \asha      &                       &                       &                      & & 0.25 & 0.24 & 0.27 && 0.16 & 0.15 & 0.20 && 0.24 & 0.23 & 0.25 \\ \hline
\multirow{3}{*}{0.10--0.15} & \acas      & \multirow{3}{*}{1995} & \multirow{3}{*}{1978} & \multirow{3}{*}{109} & & 0.06 & 0.04 & 0.05 && 0.11 & 0.09 & 0.14 && 0.04 & 0.02 & 0.03 \\
                            & \aout      &                       &                       &                      & & 0.06 & 0.04 & 0.04 && 0.14 & 0.13 & 0.16 && 0.04 & 0.02 & 0.02 \\
                            & \asha      &                       &                       &                      & & 0.28 & 0.26 & 0.29 && 0.18 & 0.16 & 0.16 && 0.26 & 0.25 & 0.27 \\ \hline
\multirow{3}{*}{0.04--0.15} & \acas      & \multirow{3}{*}{4537} & \multirow{3}{*}{4514} & \multirow{3}{*}{245} & & 0.06 & 0.04 & 0.04 && 0.10 & 0.09 & 0.11 && 0.05 & 0.02 & 0.03 \\
                            & \aout      &                       &                       &                      & & 0.07 & 0.04 & 0.03 && 0.13 & 0.12 & 0.15 && 0.05 & 0.02 & 0.02 \\
                            & \asha      &                       &                       &                      & & 0.26 & 0.25 & 0.28 && 0.17 & 0.15 & 0.19 && 0.25 & 0.24 & 0.26 \\ \hline
\enddata
\tablecomments{We randomly perturb the measured asymmetry parameters based on 
their uncertainties and calculate the averaged mean, standard deviation 
($\sigma$), and median of the samples.  The uncertainties of these parameters 
based on bootstrapping are always smaller than 0.01.
Column (1): Redshift range.
Column (2): The three asymmetry parameters.
Columns (3)--(5): The number of sources in each sample.
Columns (6)--(8): The mean asymmetries of each sample.
Columns (9)--(11): The standard deviation of the asymmetries of each sample.
Columns (12)--(14): The median asymmetries of each sample.
}
\end{deluxetable*}

\subsection{Major Merger Fraction}
\label{ssec:fmerg}

\begin{figure*}[htbp]
\centering
\includegraphics[width=\textwidth]{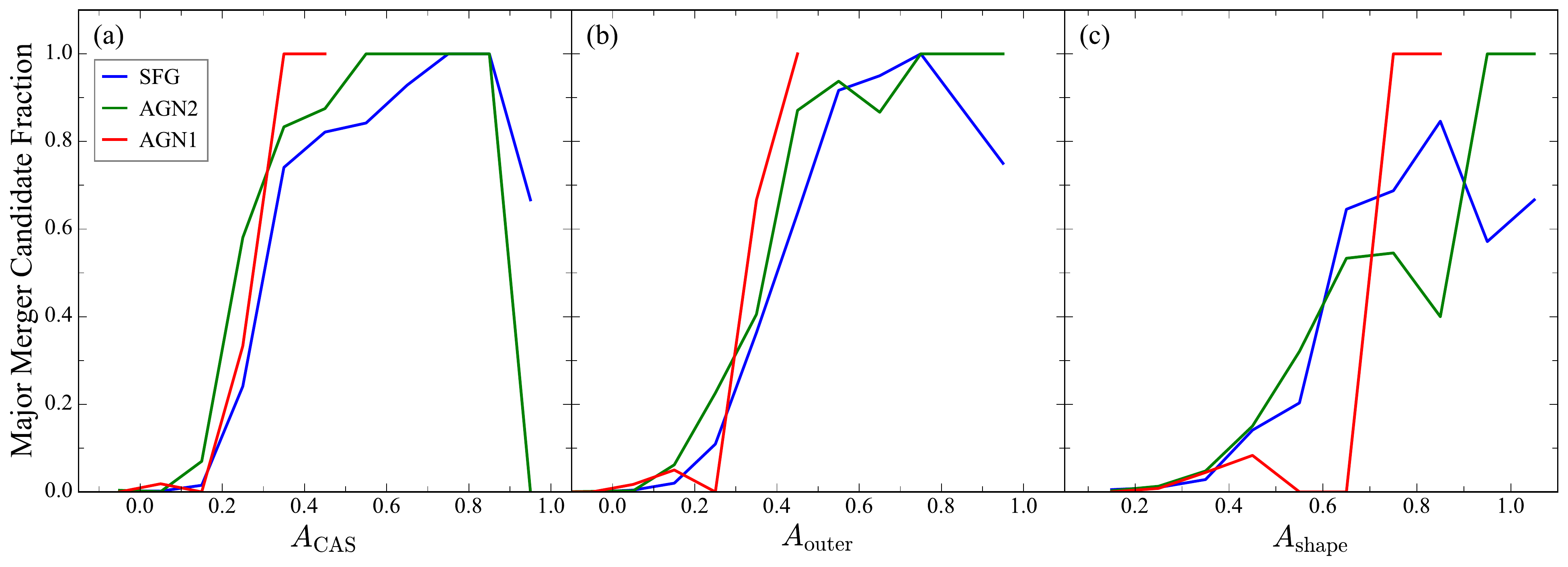}
\caption{Fraction of major merger candidates (number ratio between major merger candidates and all objects) as a function of (a) $A_{\rm CAS}$, (b) $A_{\rm outer}$, and $A_{\rm shape}$ for star-forming galaxies (SFG; blue), type~2 AGNs (AGN2; green), and type~1 AGNs (AGN1; red).}
\label{Fig:f_merger}
\end{figure*}

As discussed in Section~\ref{ssec:mask}, we assume that any extended source projected within 2~\rpet\ of the primary target having at least 25\% of its flux to be a possible companion galaxy.  We refer to these systems as major merger candidates, with the caveat that the ``post-mergers'' when the two galaxies are fully coalesced into one system are not being considered by this definition. Among our total sample of 9296 galaxies, 293 ($\sim 3\%$) are identified as major merger candidates, which are split roughly evenly among the galaxy subsamples (146/4537 SFGs, 3.2\%; 138/4514 AGN2s, 3.1\%, and 9/245 AGN1s, 3.7\%).  This overall statistic is consistent with the fraction of major mergers ($\sim 1\%-6\%$) among low-redshift galaxies with stellar mass comparable to that of our sample galaxies ($M_* \gtrsim 10^{10}\,M_\odot$; e.g., \citealt{Darg2010MNRAS,Mantha2018MNRAS,Thibert2021RNAAS}).  Figure~\ref{Fig:f_merger} illustrates that the fraction of major merger candidates quickly reaches nearly 100\% when \acas\ and \aout\ exceed $\sim 0.4$.  Although the merger fraction also increase toward higher values of \asha, this parameter is less sensitive to mergers.  In agreement with previous studies \citep{Conselice2000ApJ,Wen2014ApJ}, we concur that non-parametric asymmetry parameters offer a highly effective tool to identify galaxy mergers, although previous studies, again, usually include post-mergers which are also effectively selected using the asymmetry.  The overall consistency with previous works supports our assumption that most of the major merger candidates are indeed physically related instead of projected companions.

\subsection{Relation with Nuclear Activity}
\label{ssec:nuc}

Most AGNs in our sample are relatively symmetric.  Unless the asymmetric features of interaction quickly vanish in the $\lesssim 1~\mathrm{Gyr}$ time lag between the merger event and the activation of the AGN \citep{Wild2010MNRAS,Hopkins2012MNRAS,Blank2016MNRAS,Yesuf2020ApJ}, our results suggests that galaxy-galaxy interactions and mergers, particularly major mergers, are not the primary mechanism responsible for triggering nuclear activity in these systems.  In contrast to the evolutionary scenario, we also see no evidence that asymmetry depends on AGN type.  AGN2s have similar levels of asymmetry as AGN1s.  Neither of these results is surprising, in light of the very modest levels of AGN activity present in the local Universe.  For example, our sample of AGN2s has a median \OIII\ luminosity of $10^{40.6}\,\mathrm{erg\,s^{-1}}$, whereas for AGN1s \loiii\ $= 10^{41.0}\,\mathrm{erg\,s^{-1}}$, which, for a bolometric correction of $L_{\rm bol} \approx 600$ \loiii\ \citep{Kauffmann2009MNRAS}, corresponds to $L_{\rm bol} \approx 10^{43.4}$ and $10^{43.8}\,\mathrm{erg\,s^{-1}}$, respectively. With a median $M_* \approx 10^{10.8}\,M_\odot$, $M_{\rm BH} \approx 10^{8}\,M_\odot$, and AGN2s only reach an Eddington ratio $\sim 0.002$, and for AGN1s  $\sim 0.005$\footnote{The foregoing estimates are based on \OIII\ luminosities that have been corrected for Galactic extinction but not internal extinction.  Zhuang \& Ho (2020) report a median internal extinction of $A_V = 1.1$ mag for both low-redshift AGN1s and AGN2s selected from SDSS, not dissimilar from those studied here. Correcting for this amount of extinction would boost the luminosities and Eddington ratios by a factor of $\sim 3$, not enough to materially alter our conclusions.}.  Albeit substantially more active than the nearest supermassive BHs \citep{Ho2008ARAA}, the level of nuclear activity in our sample is still much lower than that of the bulk of low-redshift broad-line AGNs (e.g., \citealt{Greene2007ApJ,Liu2019ApJS}) and type~2 quasars \citep{Kong2018ApJ}.  These relatively low-luminosity AGNs can be sustained largely by internal sources of fueling mediated by secular processes \citep{Ho2009ApJ} without the aid of external gas supply furnished through dynamical interactions.  The merger fraction of AGNs studied by \cite{Treister2012ApJ}, which have bolometric luminosity $\sim 10^{43.5}\,\mathrm{erg\,s^{-1}}$, agrees well with our results.  Major mergers only matter for more powerful sources (e.g., \citealt{Weigel2018MNRAS,Kim2021ApJS}). For instance, studies based on images taken with the Hubble Space Telescope find a major merger fraction of $\sim 20\%$ for type~1 quasars \citep{Kim2017ApJS,Zhao2021ApJ} and 
$\sim 40\%$ for type~2 quasars \citep{Zhao2019ApJ}.

\begin{figure}[htbp]
\centering
\includegraphics[width=0.5\textwidth]{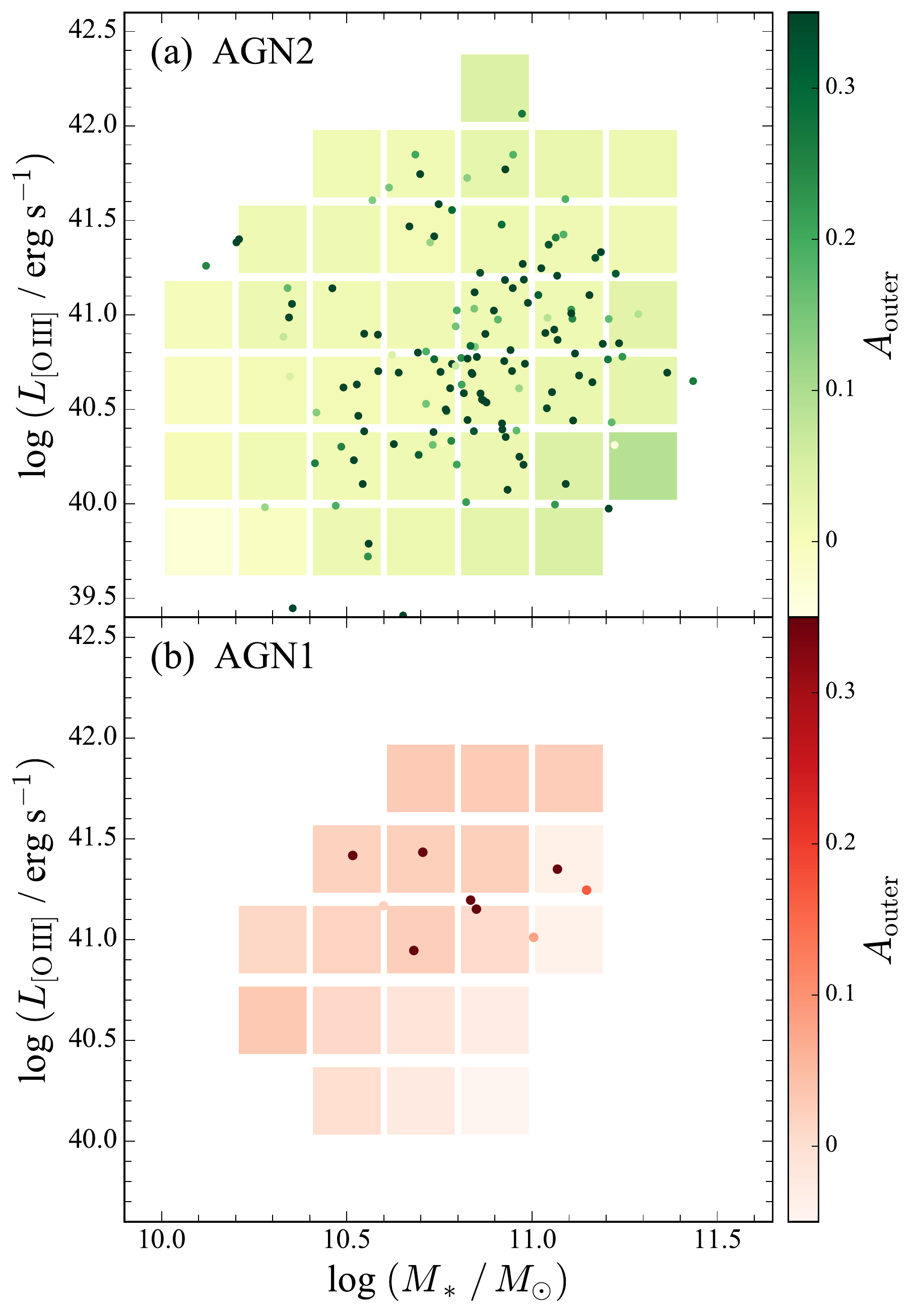}
\caption{The distribution of \loiii\ vs. $M_*$ for (a) type~2 AGNs and (b) type~1 AGNs, color-coded by the \aout\ asymmetry index.  Small circles are major merger candidates (Section~\ref{ssec:fmerg}), and large squares represent the median values of the other objects in bins of $M_*$ and \loiii.  We require that there are $\ge10$ objects per bin in (a) and $\ge5$ objects per bin in (b).}
\label{Fig:Ms_OIII_A_CAS}
\end{figure}

Does the degree of asymmetry correlate with the level of BH accretion in our sample?  We investigate this issue by examining the distribution of asymmetry as a function of \loiii\ and $M_*$, focusing solely on the \aout\ parameter, for which we have the greatest confidence across all three galaxy samples (Section~\ref{sec4:asy}; Appendix A).  We separately identify the major merger candidates (Section~\ref{ssec:fmerg}), whose asymmetry parameter was measured for the galaxy pair while the \loiii\ and $M_*$ pertain to the single central galaxy.  Figure~\ref{Fig:Ms_OIII_A_CAS} indicates that \aout\ is distributed roughly uniformly for both AGN1s and AGN2s.  On average, \aout\ among AGN2s shows a very moderate tendency to increase toward higher $M_*$, which likely reflects the environmental dependence of stellar mass \citep{Hopkins2010ApJ,Weigel2018MNRAS}.  We hesitate to draw firm conclusions about the AGN1s, given their limited statistics. The targets identified as major merger candidates show no preference to host more luminous AGNs.  Although \acas\ is less robust for AGN1s, we confirm that all of the conclusions in Figure~\ref{Fig:Ms_OIII_A_CAS} hold if we replace \aout\ with \acas.  Our results resonate with those of \cite{Stemo2021}, whose analysis of a large sample of dual AGNs reveals that the AGN luminosity of mergers is not significantly different from that of the overall AGN population, while standing in contrast to the general expectation that BH accretion should be enhanced by galaxy interactions \citep{Ellison2011MNRAS,VanWassenhove2012ApJ,Satyapal2014MNRAS}.

\subsection{Relation with Star Formation}
\label{ssec:sfr}

\begin{figure*}
\centering
\includegraphics[width=1.0\textwidth]{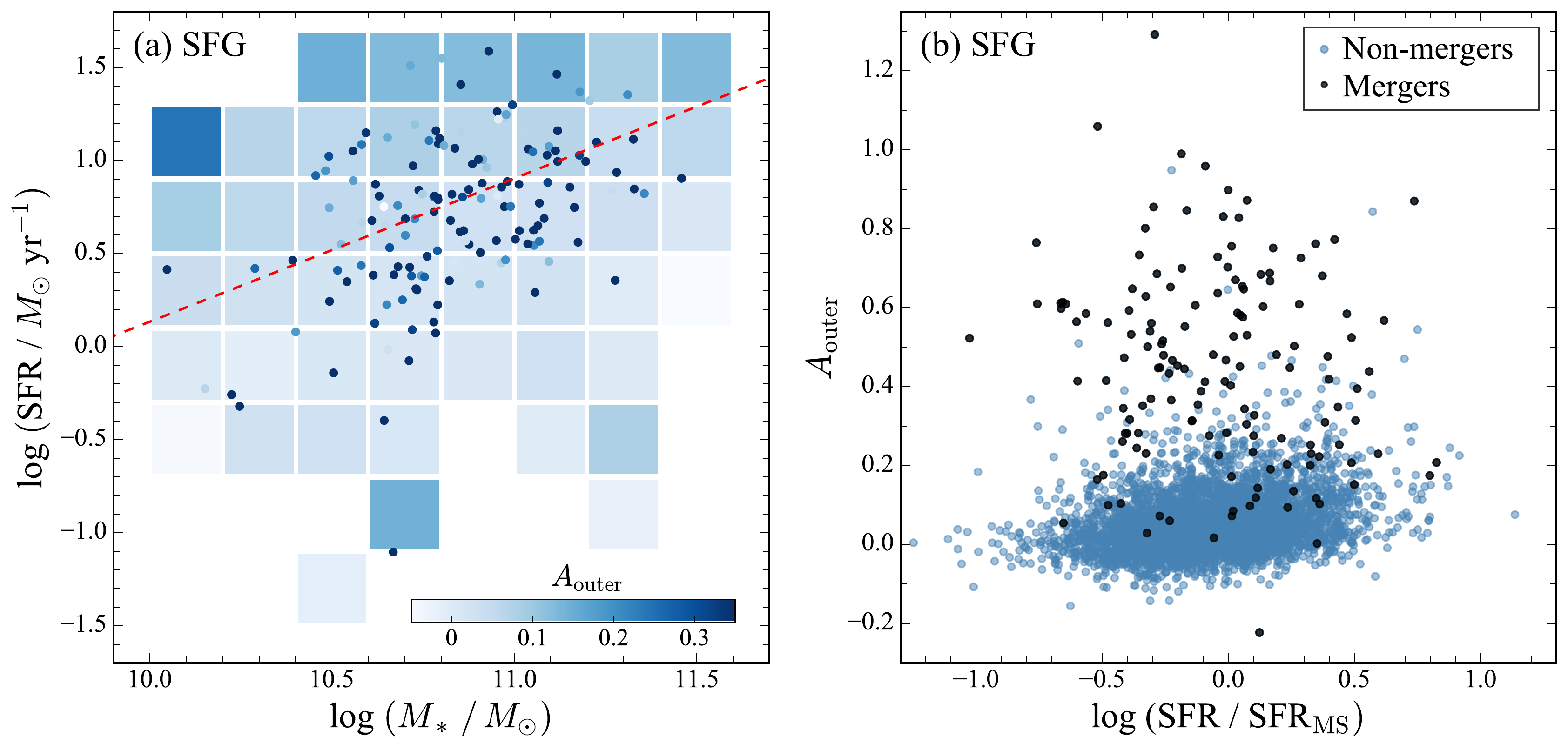}
\caption{(a) The distribution of SFR vs. $M_*$ for the SFG sample, color-coded by the \aout\ asymmetry index.  Small circles are major merger candidates (Section~\ref{ssec:fmerg}), and large squares represent the median values of the other objects in bins of $M_*$ and SFR.  We require that there are $\ge 10$ objects per bin. The dashed line is the local SFG main sequence from \cite{Elbaz2007AA}, converted to the stellar initial mass function of \cite{Chabrier2003PASP}.  (b) The relation between \aout\ asymmetry and the distance from the main sequence, $\log\,({\rm SFR}/{\rm SFR}_{\rm MS})$, where SFR$_{\rm MS}$ is the SFR along the main sequence.}
\label{Fig:MS_A_CAS}
\end{figure*}

Although not the main focus of this study, we briefly mention, for completeness, the relation between morphological asymmetry and star formation, as it pertains to the distribution of SFGs on the galaxy main sequence (e.g., \citealt{Daddi2007ApJ,Elbaz2007AA,Noeske2007ApJ}).  It is evident from Figure~\ref{Fig:MS_A_CAS} that the average level of asymmetry (\aout) increases mildly but systematically above the main sequence, echoing the recent findings of \cite{Yesuf2021}.  Consistent with previous work \citep[e.g.,][]{Cibinel2019MNRAS}, major merger candidates (Section~\ref{ssec:fmerg}) stand out the most, but it is notable that a significant fraction of major merger candidates also lie on and below the main sequence (Figure~\ref{Fig:MS_A_CAS}b).  From their morphological classication of over 200,000 galaxies using convolutional neutral networks, \cite{Pearson2019AA} recently also reported that mergers do not show significantly different SFRs compared to non-mergers, although they find that the fraction of mergers is significantly higher among galaxies above the main sequence.  Perhaps the major merger candidates below the main sequence are dominated by projected pairs and are really not interacting systems.  We visually checked these targets but could not see any obvious clues.  Tidal features are clearly seen in some of them, proving that they are true physical mergers. Another possibility is that the low-SFR pairs signify dry mergers or mixed mergers, which dominate the merger population  at low redshifts \citep{Lin2008ApJ}, wherein at least one of the member galaxies is cold gas-deficient.

\section{Summary}
\label{sec7:sum}

The morphological asymmetry parameter offers an efficient, model-independent measure to constrain the local environment of large samples of galaxies.  This work uses three asymmetry indicators (\acas, \aout, and \asha), conveniently built into the Python package \statm, to quantify the morphological asymmetry of the $i$-band images of over 9000 nearby ($0.04<z<0.15$), massive ($M_*>10^{10}\,M_\odot$) galaxies drawn from the SDSS spectroscopic database and the Pan-STARRS1 $3\pi$ Steradian Survey, to investigate the role of mergers and interactions in triggering nuclear activity. The active galaxies, comprising both type~1 and type~2 AGNs, are carefully matched to a control sample of star-forming galaxies.  We develop a comprehensive strategy to effectively isolate the signal from the target galaxy and associated neighbors sizable enough to be a candidate major merger, all the while masking other contaminating galaxies and field stars.  We perform a variety of experiments to test the efficacy and limitations of the three asymmetry indicators, as well as to obtain robust estimates of their uncertainties.

Our main results are as follows: 

\begin{itemize}

\item While all three indicators can trace visible morphological distortions over a wide dynamic range in asymmetry for the chosen sample and image quality of the Pan-STARRS survey, the parameters
\acas\ and \aout\ perform best for galaxies without a prominent nucleus (star-forming galaxies and type~2 AGNs), \aout\ can be effective even in the presence of a bright nucleus (type~1 AGNs), and \asha\ should be avoided for all galaxy types because of its sensitivity to background noise.

\item Roughly 3\% of the galaxies at $z<0.15$ are likely major merger systems that can be identified effectively as having \acas\ and \aout\ values $\gtrsim 0.4$.
 
\item Contrary to the expectations of the merger-driven scenario for AGN evolution, the host galaxies of AGNs exhibit lower degrees of asymmetry than the control sample of star-forming galaxies.  Both type~1 and type~2 AGNs also have comparable levels of asymmetry.  We find no clear correlation between asymmetry and AGN luminosity, not even for major merger candidates.

\item A large portion of the star-forming galaxies with the largest asymmetry lie offset above the star-forming main sequence in the regime occupied by starbursts, but not all major merger candidates are actively forming stars, possibly because they are gas-poor.

\end{itemize}

\begin{acknowledgments}
Y. Z. thanks Ruancun Li for his helpful technical advice. This research was supported by the National Science Foundation of China (11721303, 11991052) and the National Key R\&D Program of China (2016YFA0400702). 
\end{acknowledgments}


\appendix

\section{Uncertainties of Asymmetry Parameters}
\label{apd:unc}

We follow the method of \cite{Shi2009ApJ} to calculate the uncertainty of the CAS asymmetry, and we generalize the method to calculate the uncertainty of the other asymmetry estimators.  We show that the typical uncertainty of individual targets is consistent with the width of the negative part of the measured sample distribution, indicating that noise from the background dominates the uncertainties of the CAS and outer asymmetry.  Although the shape asymmetry is not affected by the background, we can adopt a similar method to estimate its uncertainty using the measured sample distribution of asymmetries.

Based on mock images generated from observed galaxies and the background without source emission, \cite{Shi2009ApJ} perform tests to estimate the background CAS asymmetry (see Equation 1) and conclude that the minimum value of the background asymmetry used by \cite{Conselice2000ApJ} overestimates the total asymmetry.  Shi et al. measure the background asymmetry by randomly sampling regions around the target galaxy, estimating it with the 15th percentile of the sampled distribution, because the correction of the background noise tends to underestimate the asymmetry of the source.  They find that the real $3\,\sigma$ uncertainty of the background asymmetry is $\sim 2$ times the RMS of the sampled background asymmetry. The latter is adopted as a conservative estimate of the uncertainty of the CAS asymmetry.

\begin{figure}
\figurenum{A1}
\centering
\includegraphics[width=0.45\textwidth]{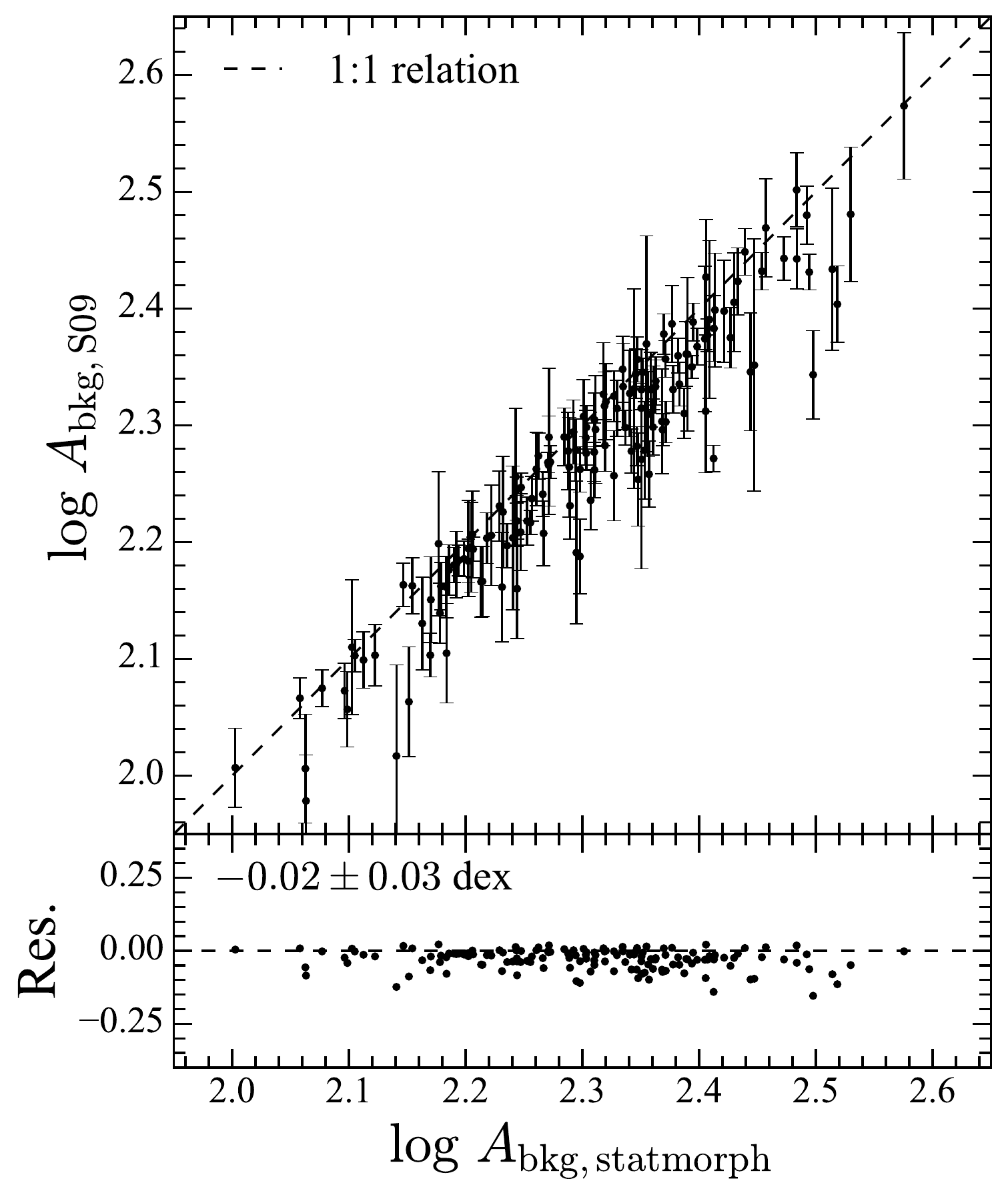}
\caption{Comparison of the background asymmetry measured by the original \statm\ method with our method of random sampling, which follows \cite{Shi2009ApJ}.  For clarity, only 150 randomly selected targets are plotted; the results for the entire sample are statistically the same. The median and RMS of the deviation between the two methods is given on the upper-left corner of the bottom panel, which shows the residuals of the one-to-one relation (dashed line) in the top panel.}
\label{fig:unc1}
\end{figure}

Defined as $\abkg=\Sigma |B_0-B_{180}| / N_\mathrm{pixels}$, the background asymmetry originally estimated in \statm\ is calculated simply from the asymmetry of a background region that is free of any masked pixels.  We adopt this definition but measure \abkg\ by randomly sampling the image.  We require the sampled region to be masked by no more than 10\% and only use the unmasked pixels to calculate \abkg.  Following \cite{Shi2009ApJ}, we then adopt the 15 percentile of the sampled distribution as the estimated background asymmetry.  As shown in Figure~\ref{fig:unc1}, our values of the background asymmetry, denoted by $A_\mathrm{bkg,S09}$, are slightly lower than, but still fall within the $1\,\sigma$ scatter of, the values obtained through \statm\ ($A_\mathrm{bkg,statmorph}$).  This is expected because $A_\mathrm{bkg,statmorph}$ is on average close to the median (50 percentile) of the randomly sampled \abkg.  \cite{Thorp2021MNRAS} recently also use simulated galaxies to study how to properly correct the background asymmetry.  Calculating the background asymmetry from a region offset from the source using \statm, they find that the intrinsic asymmetry can be recovered best by adopting the measured background asymmetry after scaling it down by $\sim$15\%.

We adopt 2 times the RMS of the randomly sampled \abkg\ as the estimated uncertainty of \abkg\ of the source asymmetry.  We propagate the uncertainty of \abkg\ according to Equation~(\ref{eq:Asymmetry}) and obtain the ($3\,\sigma$) uncertainty of the source asymmetry.  This method was proposed by \cite{Shi2009ApJ} for the CAS asymmetry.  We adopt the same approach to estimate the uncertainty of the outer asymmetry and confirm its robustness below.  The typical uncertainties of \acas\ and \aout\ of the SFG sample are 0.03 and 0.06, respectively, with an RMS of $\sim 0.03$.  The uncertainty of \aout\ is somewhat larger because the total flux (the denominator in Equation 1) used for \aout\ is usually lower than that for \acas.  

\begin{figure*}
\figurenum{A2}
\centering
\includegraphics[width=0.9\textwidth]{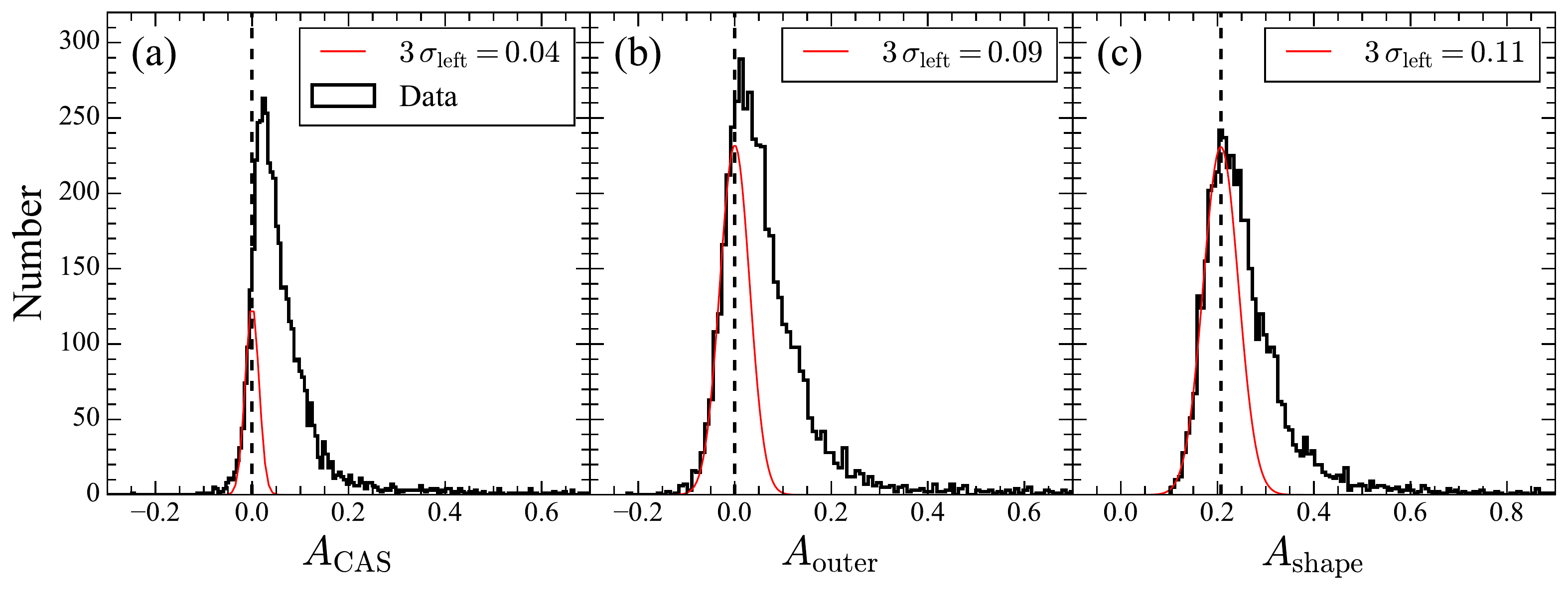}
\caption{Distribution of the measured (a) \acas, (b) \aout, and (c) \asha\ of the SFG sample (black histogram).  A Gaussian model (red curve) is fitted to the part of each distribution to the left of the dashed vertical line.  The $3\,\sigma$ of the best-fit Gaussian model is displayed in the legend of each panel.}
\label{fig:unc_sf}
\end{figure*}

\cite{Shi2009ApJ} studied images of galaxies observed with the Hubble Space Telescope, which have very different characteristics compared to the ground-based images from Pan-STARRS investigated here.  In order to test the uncertainty estimates of our galaxies derived using Shi et al.'s method, we compare the sample distribution of \acas\ and \aout\ with the typical uncertainties for individual targets.  First, consider the SFG sample.  The distributions of the asymmetry parameters display a long tail toward higher value (Figure~\ref{fig:unc_sf}).  Since the asymmetry parameters cannot be lower than zero if they are noise-free, we can use the negative part of the sample distribution to estimate the average uncertainty of \acas\ and \aout.  We fit a Gaussian profile to the negative part of \acas\ and \aout\ distributions, fixing the mean to 0.  The $3\,\sigma$ value of the best-fit Gaussian (red curve in Figure~\ref{fig:unc_sf}) is 0.04 for \acas\ and 0.09 for \aout, which is largely consistent with but slightly larger than the typical uncertainties of individual targets reported above.  This confirms that the uncertainty introduced by the background primarily contributes to the error budget of our \acas\ and \aout\ measurements of the SFGs.  Previous studies \citep[e.g.,][]{Conselice2000ApJ,Thorp2021MNRAS} found that the asymmetry is dominated by the background noise if the source S/N is below 100.  With a typical S/N $\gtrsim 100$, our galaxies should have reliable asymmetry measurements.  The asymmetry center of the target is well constrained, and we confirm that the uncertainty of the asymmetry is dominated by the noise of the background and is inversely proportional to the source flux.

\begin{figure*}
\figurenum{A3}
\centering
\includegraphics[width=0.9\textwidth]{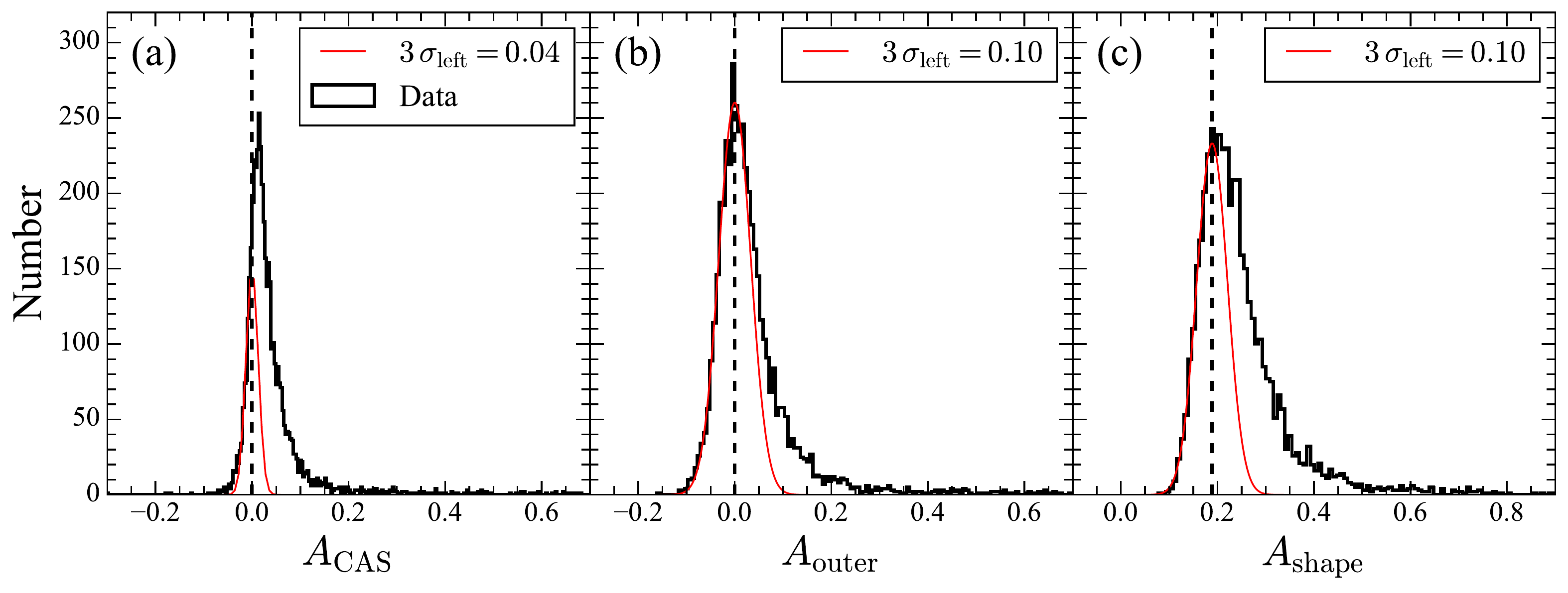}
\caption{Distribution of the measured (a) \acas, (b) \aout, and (c) \asha\ of the AGN2 sample (black histogram).  A Gaussian model (red curve) is fitted to the part of each distribution to the left of the dashed vertical line.  The $3\,\sigma$ of the best-fit Gaussian model is displayed in the legend of each panel.}
\label{fig:unc_t2}
\end{figure*}

\begin{figure*}
\figurenum{A4}
\centering
\includegraphics[width=0.9\textwidth]{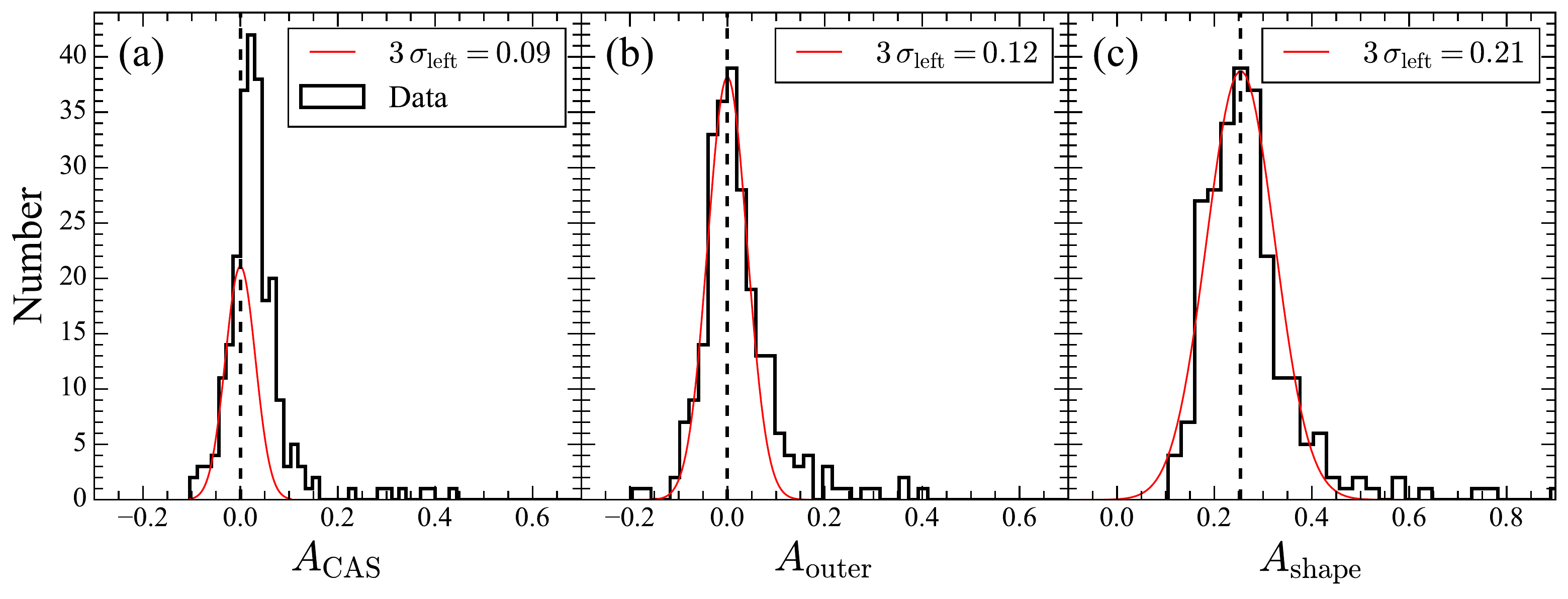}
\caption{Distribution of the measured (a) \acas, (b) \aout, and (c) \asha\ of the AGN1 sample (black histogram).  A Gaussian model (red curve) is fitted to the part of each distribution to the left of the dashed vertical line.  The $3\,\sigma$ of the best-fit Gaussian model is displayed in the legend of each panel.}
\label{fig:unc_t1}
\end{figure*}

We confirmed that the method proposed by \cite{Shi2009ApJ} can be applied to Pan-STARRS images of SFGs to estimate the uncertainties of \acas\ and \aout.  Does the same conclusion hold for AGNs?  The typical uncertainties of \acas\ and \aout\ of both the AGN2 and AGN1 samples, based on Shi et al.'s method, are very close to those of the SFGs.  The uncertainties estimated from the AGN2 sample distributions (Figure~\ref{fig:unc_t2}) are also very close to those of the SFG sample.  This is not surprising because AGN2s, lacking bright nuclei, are not dissimilar from SFGs from a measurement point of view.  However, the situation for AGN1s appears somewhat more complicated (Figure~\ref{fig:unc_t1}).  The sample distribution of \acas\ for AGN1s indicates an uncertainty that is more than twice as large as that derived from Shi et al.'s method.  This is likely due to the contamination by the nuclear emission.  The largely symmetric PSF will reduce the total asymmetry, such that AGN1s would appear to have lower intrinsic \acas, whose distribution at $\acas<0$ is broader than those of SFGs and AGN2s.  On the other hand, the uncertainties of \aout\ estimated from the sample distribution are still largely consistent with those derived from Shi et al.'s method, because the asymmetry is only calculated in the outer region of the host galaxy.  We conclude that our uncertainties estimated from Shi et al.'s method are robust for both \acas\ and \aout\ when applied to AGN2s, and for \aout\ when applied to AGN1s.  The \acas\ asymmetry of AGN1s likely is affected significantly by the nuclear emission and therefore should not be used.

Since the shape asymmetry does not require a correction for the background asymmetry, we estimate its uncertainty from the measured sample distribution of the asymmetry parameter.  Fitting a Gaussian model to the measured \asha\ distribution on the left side of the peak, which is centered at $\asha \approx 0.2$ (Figures~A1c--A3c), the $3\,\sigma$ width of SFGs, AGN2s, and AGN1s are 0.11, 0.10, and 0.21, respectively.  The shape asymmetry, being more sensitive to the lower S/N features of the galaxy \citep{Pawlik2016MNRAS}, is expected to show lightly larger uncertainty than \acas\ and \aout.  As discussed in Section~\ref{sec4:asy}, we will not rely on \asha\ for our main analysis.

\end{document}